\documentclass[11pt]{article}

\usepackage[margin=1in]{geometry}
\usepackage[utf8]{inputenc}
\usepackage[T1]{fontenc}
\usepackage{microtype}
\usepackage{lmodern}
\usepackage{amsmath,amssymb}
\usepackage{graphicx}
\usepackage{booktabs}
\usepackage{multirow}
\usepackage{array}
\usepackage{xcolor}
\usepackage{caption}
\usepackage{subcaption}
\usepackage{placeins}
\usepackage{xspace}
\usepackage[hidelinks,
            pdftitle={FCDC: Nonvolatile Charge-Domain Attention with HZO Ferroelectric Capacitors},
            pdfauthor={Fares Abouagor},
            pdfsubject={Hardware Architecture; Compute-in-Memory; Large Language Models},
            pdfkeywords={ferroelectric, HZO, compute-in-memory, charge-domain, transformer, attention, KV cache},
            pdfcreator={pdflatex},
            pdfproducer={pdflatex}]{hyperref}

\newcommand{\fcdc}{FCDC\xspace}

\title{FCDC: Nonvolatile Charge-Domain Attention\\with HZO Ferroelectric Capacitors}
\author{Fares Abouagor \\
  \small Faculty of Engineering, Mansoura University, Egypt \\
  \small \texttt{faresaboagour@std.mans.edu.eg}}
\date{June 2026}

\begin{document}
\maketitle
\thispagestyle{empty}
\begin{center}\small\itshape Preprint. Code: \url{https://github.com/faris-agour/FCDC}\end{center}

\begin{abstract}
Transformer decoding is increasingly constrained by the key-value (KV) cache that must stay resident and be re-read for the life of a session. We present the \emph{Ferroelectric Charge-Domain Compute Cell} (\fcdc), a hafnium-zirconium-oxide (HZO) memcapacitor that stores analog weights as nonvolatile remanent polarization and performs charge-domain vector--matrix multiplication for attention. A full-substrate mode (all $q,k,v,o$ projections and both attention matmuls on \fcdc) provides the harder noise test and upper-bounds the narrower KV-coprocessor serving mode (KV storage plus the two matmuls).

The evaluation is simulation-based (no \fcdc device is fabricated), cross-checked across four simulators, and anchored to wafer-scale 10\,nm-HZO measurements \cite{anu2026wafer, li2024endurance}. Across 12 pretrained LLMs (dense to Qwen3-32B fully substituted, with a 141\,B Mixtral-8$\times$22B partial-layer stress test), all-layer noise substitution adds $+2.6\%$ WikiText-2 perplexity on Qwen3-32B and $+2.9\%$ (five-seed mean) on Mistral-7B-v0.3; five downstream tasks stay within $5\%$ of digital, the deltas hold to $128$\,k context, and the serving mode costs under $0.5\%$ at $7$--$8$\,B. Analog-input fragility localizes to the value projection, and periphery-side input dithering recovers the worst-case PWM-nonlinearity collapse to near-baseline without retraining.

The advantage is not raw multiply-accumulate energy, where the \fcdc tile merely matches switched-capacitor SRAM compute-in-memory. It is nonvolatility, no refresh, and KV-cache residency. On measured INT4 decode energy, a workload simulator projects $18$--$35\times$ lower per-served-token energy on retrieval-augmented generation and agent loops than a single-user GPU, narrowing to $1.4$--$4.7\times$ against optimized serving baselines (batched vLLM, CPU+NVMe parking, power-gating) but exceeding $40\times$ on multi-hour parked sessions. Long-residency, persistent-KV serving is the regime where a nonvolatile charge-domain substrate holds a durable advantage over an optimized GPU.
\end{abstract}

\section{Introduction}

Autoregressive transformer inference repeatedly evaluates attention over a growing KV cache. The cost is partly arithmetic, through $Q\!\cdot\!K^\top$ and $\mathrm{softmax}(\cdot)\!\cdot\!V$, and partly data movement, because stored keys and values must remain accessible throughout decoding. This makes attention a natural target for in-memory compute, but it also imposes a stronger requirement than conventional weight-stationary accelerators: the stored state is dynamic, long-lived, and repeatedly read.

Existing device approaches address different parts of this problem. Gain-cell analog attention performs attention multiplications inside a charge-storage array, but the storage is volatile and requires refresh on millisecond timescales \cite{leroux2025}. Ferroelectric KV-cache arrays demonstrate nonvolatile storage with fast switching and high endurance in specific 3D stacks, but the ferroelectric array is primarily a storage substrate rather than the attention compute engine \cite{xu2025}. Ferroelectric field-effect transistor (FeFET) content-addressable-memory (CAM) and compute-in-memory (CIM) work further shows that ferroelectric devices can support KV-cache pruning and in-memory search operations. These results motivate a narrower question: can a capacitive HZO ferroelectric cell provide a useful design point for nonvolatile KV-cache residency and local charge-domain attention compute?

This paper presents \fcdc, a modeled HZO memcapacitor tile for charge-domain vector--matrix multiplication (VMM). The tile stores analog state in a ferroelectric capacitor, reads it through an access device, and accumulates column charge for attention projections and attention matrix products. A negative-capacitance read path is analyzed as an optional way to improve signal margin.

The contribution is a device-to-system evaluation built on cross-simulator consistency checks and on recent wafer-scale HZO capacitor data. At the circuit and array level, the cell noise and peripheral energy budgets are derived and the same analytic model is cross-checked across \textsc{ngspice}, \textsc{CrossSim}, \textsc{FiPy}, and \textsc{NeuroSim}. At the model level, the resulting noise impact on pretrained LLMs is measured at scale, including projection-only substitution, end-to-end analog attention, seed sensitivity, longer-context replication, and a small low-rank-adapter quantization-aware training (LoRA QAT) recovery path. A per-projection resilience study then localizes analog-input fragility to the value projection and shows that periphery-side input dithering neutralizes the binding PWM input non-ideality without retraining (\S\ref{sec:stride}). At the system level, the active multiply-accumulate (MAC) energy with full peripheral accounting is compared against an analytic NVIDIA A40 baseline and against measured analog in-memory-compute (IMC) silicon, and a workload-level simulator quantifies per-served-token energy across five serving regimes. Throughout, simulation-backed results are separated explicitly from the fabrication-dependent assumptions they rest on (Table~\ref{tab:evidence}).

\section{Related Work}\label{sec:related}

\begin{table}[htbp]
\centering
\footnotesize
\setlength{\tabcolsep}{4pt}
\caption{Position relative to the closest prior work on nonvolatile / charge-domain attention substrates. ``In-place'' = matrix multiplication executed inside the storage array. ``Dense'' = full attention without top-$k$ pruning.}
\label{tab:position}
\begin{tabular}{@{}p{0.15\linewidth}p{0.18\linewidth}p{0.13\linewidth}p{0.14\linewidth}p{0.18\linewidth}p{0.10\linewidth}@{}}
\toprule
Work & Device & Nonvolatile & In-place & LLM evaluated & Context \\
\midrule
Leroux \cite{leroux2025}      & gain cell           & no (ms refresh) & yes (dense)    & GPT-2 (124\,M)  & $\sim$1\,k \\
Xu \cite{xu2025}              & 3D 1T-nC-1T FE      & yes             & pruning+hybrid & device paper    & --        \\
UniCAIM \cite{unicaim2025}    & FeFET CAM/CIM       & yes             & top-$k$        & GPT-2 class     & 2\,k      \\
Yin \cite{yin2024onefefet}    & 1FeFET-1C           & yes             & yes (macro)    & macro-level     & --        \\
\textbf{\fcdc (ours)}         & \textbf{HZO 1T-1C}  & \textbf{yes}    & \textbf{yes (dense)} & \textbf{Qwen3-32B}        & \textbf{$\le$128\,k}$^\ast$ \\
\bottomrule
\end{tabular}
\\[2pt]
{\scriptsize $^\ast$Largest dense model fully substituted: Qwen3-32B at 4\,k tokens. Longest context replicated: Mistral-7B-v0.3 at 128\,k tokens. Mixtral-8x22B (141\,B mixture-of-experts, MoE) reported at $k{=}75\%$ partial-layer only.}
\end{table}

\paragraph{Analog in-memory attention.} Leroux et al. \cite{leroux2025} perform $Q\!\cdot\!K^\top$ and $\mathrm{softmax}(\cdot)\!\cdot\!V$ inside a gain-cell array and report GPT-2-comparable text quality at the 124\,M scale on OpenWebText / WikiText-2 / LAMBADA. Their array is volatile and must be refreshed on millisecond timescales (the silicon CMOS gain-cell retention constant reported in that work is $\tau\!\approx\!5$\,ms). The main distinction of \fcdc is therefore nonvolatile KV-cache residency rather than the idea of analog attention itself.

\paragraph{Ferroelectric and nonvolatile KV-cache accelerators.} Xu et al. \cite{xu2025} demonstrate a 3-D vertical 1T-nC-1T ferroelectric KV-cache array with fast switching, long retention, high endurance, hybrid analog-digital CIM, and token-wise pruning. UniCAIM \cite{unicaim2025} is even closer at the architecture level: it uses FeFET-based CAM/CIM for long-context LLM KV-cache pruning, charge-domain score accumulation, and current-domain exact attention on selected tokens. \fcdc differs from both in adopting a minimal capacitor-based (1T-1C) cell and in evaluating it as a dense, nonvolatile resident-KV attention substrate under explicit peripheral and device-physics bounds.

\paragraph{Ferroelectric charge-domain CIM.} FeFET-based nonvolatile charge-domain CIM was introduced before this work \cite{yin2021cdfefet}, and the 1FeFET-1C neuro-symbolic AI prototype \cite{yin2024onefefet} demonstrates fabricated ferroelectric charge-domain MAC and associative-search functionality. Recent nonvolatile-capacitor work further studies ferroelectric capacitive crossbars and calibrated SPICE models for charge-domain MACs \cite{vadlamani2025nvcap}. These papers are the appropriate hardware baselines for the cell and macro, while HERMES \cite{legallo2023hermes} and switched-capacitor SRAM CIM \cite{verma2021charge,verma2024scsram} are the appropriate measured active-MAC energy baselines.

\paragraph{This work.} Relative to this prior art, the contribution is an LLM-tolerance evaluation of a nonvolatile HZO charge-domain attention substrate at up to 141\,B-parameter scale and 128\,k context, backed by cross-simulator consistency checks and full peripheral energy accounting. Prior ferroelectric CIM, charge-domain CIM, and ferroelectric KV-cache designs are the hardware baselines throughout, and negative-capacitance (NC) voltage gain is treated as an optional read-path technique.

\section{The FCDC Cell}\label{sec:device}

\subsection{Modeling assumptions}\label{sec:evidence}
Table~\ref{tab:evidence} summarizes the principal assumptions behind the
device-to-system model and their provenance. No \fcdc array is fabricated
in this work: device parameters are anchored to published HZO
measurements, tile and noise figures come from analytic models with a
behavioral SPICE cross-check, and energy figures are projected from the
tile model. The LLM-accuracy results are measured directly on pretrained
models, and the serving-energy analysis targets long-residency,
persistent-KV workloads.

\begin{table}[htbp]
\centering
\footnotesize
\setlength{\tabcolsep}{4pt}
\caption{Summary of modeling assumptions and their provenance. ``Literature''
= anchored to measured HZO data (not this work); ``modeled'' = closed-form
from geometry/physics; ``projected'' = from the analytic tile model;
``assumed'' = engineering estimate; ``measured'' = run on real models/GPUs
here.}
\label{tab:evidence}
\begin{tabular}{@{}>{\raggedright\arraybackslash}p{0.30\linewidth}>{\raggedright\arraybackslash}p{0.34\linewidth}>{\raggedright\arraybackslash}p{0.30\linewidth}@{}}
\toprule
Quantity & Value / setting & Provenance \\
\midrule
$P_r$, $E_c$ (HZO) & $25\,\mu$C/cm$^2$, $1$\,MV/cm & literature (10\,nm measured \cite{anu2026wafer,li2024endurance}) \\
Retention / endurance & $\geq10$\,yr / $10^{16}$ cyc & literature (3D stacks); \emph{not} transferred to 50\,nm planar \\
$C_0$ & $\approx 69$\,aF & modeled (geometry) \\
Read disturb & $\lesssim10^{-17}$/read & modeled (Merz/NLS); invalid under NC field gain \\
Noise fraction $\mathrm{nf}$ & $0.015$ nominal & analytic budget + behavioral SPICE check \\
NC read-path gain & $2.5\times$ (optional) & exploratory; needs non-hysteretic stack \\
Multilevel analog write & write-verify; $10$--$10^3\times$ sweep & modeled (Merz/NLS MC: $13$--$18\times$) \\
Operating resolution & 8-bit DAC/ADC & \emph{measured} (4-bit catastrophic, \S\ref{sec:e2e}) \\
Tile energy (PWM) & $0.94$\,fJ/MAC & projected tile model \\
LLM accuracy deltas & $+2.6$--$3.5\%$ PPL & measured (real models, \S\ref{sec:results}) \\
GPU decode energy & measured INT4 (NVML) & measured; idle $70$\,W assumed \\
Serving advantage & long-residency KV only & projected workload simulator \\
\bottomrule
\end{tabular}
\end{table}

\subsection{Device}\label{sec:cell}

Each \fcdc cell combines a nonvolatile HZO storage capacitor, an optional read-path gain element, and an access transistor.

\begin{figure}[tbp]
\centering
\begin{minipage}[c]{0.40\linewidth}
  \centering
  \includegraphics[width=\linewidth]{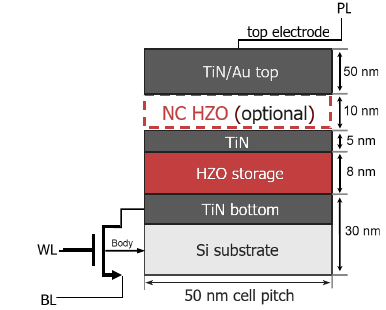}
  \subcaption{1T-1C cell cross-section.}
  \label{fig:cell}
\end{minipage}\hfill
\begin{minipage}[c]{0.57\linewidth}
  \centering
  \includegraphics[width=\linewidth]{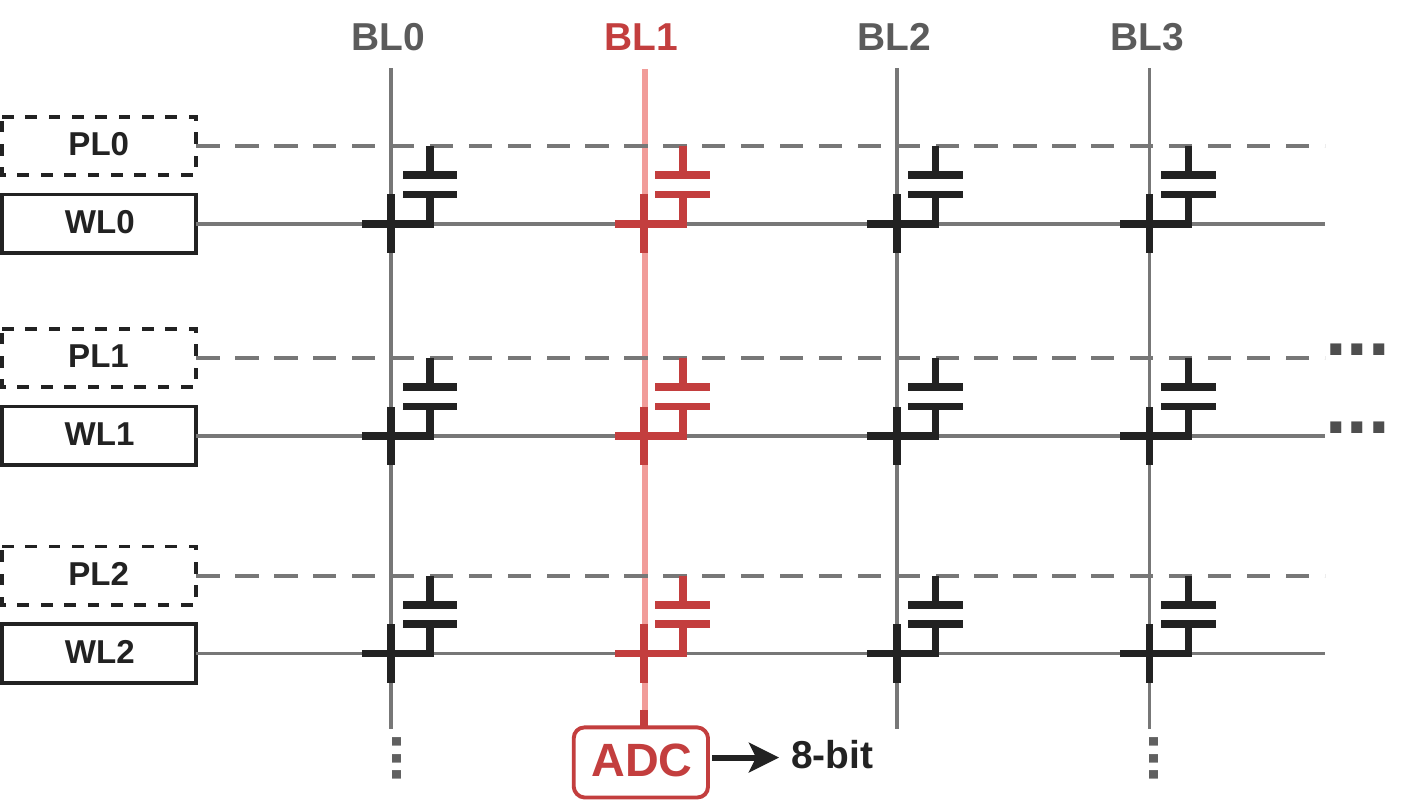}
  \subcaption{$256\!\times\!256$ tile (shown $4\!\times\!3$).}
  \label{fig:tile}
\end{minipage}
\caption{\fcdc cell and tile. (a) Storage node (TiN bottom electrode)
tied to the NMOS drain; PL drives the top electrode; optional 10\,nm
NC-HZO layer (dashed) provides read-path gain; cell pitch 50\,nm. (b)
WL selects a row, PL drives the capacitor top plates, and each column
shares one BL terminated by an 8-bit ADC; accent column shows the
selected analog read path.}
\end{figure}

\paragraph{Storage capacitor.} The storage element is a hafnium-zirconia ($\mathrm{Hf_{0.5}Zr_{0.5}O_2}$) memcapacitor storing one analog weight as remanent polarization. Capacitance
$C_0 \approx 69$\,aF at the modeled cell footprint; HZO carries the
non-volatile state with retention reported up to $\geq\!10$ years in
specific 3D 1T-nC-1T stacks \cite{xu2025} and dielectric-class
$\epsilon_r{\approx}20$--$35$ measured on free-standing HZO membranes
and related HZO stacks \cite{muller2012hzo,hzo_smallsignal_2025}. Recent small-signal
studies further decompose the apparent capacitance into dielectric +
polarization + domain-wall contributions \cite{hzo_smallsignal_2025},
which means the smooth $C(P)$ surrogate used here is an ML-grade differentiable approximation, not a device-physics model.

\paragraph{Material parameters.} The material parameters used in
this paper are within the envelope of recent measured 10\,nm HZO
capacitors. A wafer-scale study of 270 atomic-layer-deposited (ALD)
10\,nm-HZO / 30\,nm-TiN metal--insulator--metal (MIM) capacitors across six dies reports a mean
remanent polarization $\overline{P_r}\approx 40.58\,\mu\mathrm{C/cm^2}$
extracted from $\pm 6$\,V triangular dynamic-hysteresis P--V loops at
1\,kHz, with die-to-die spread captured by an unsupervised PCA / K-means
clustering model \cite{anu2026wafer}.
A separate metal--ferroelectric--insulator--metal (MFIM) HZO study at the
same 10\,nm thickness, measured by positive-up negative-down (PUND)
pulsing at $\pm 3$\,V / 100\,kHz, demonstrates
$>10^{9}$ programming/erase cycles at room temperature with recoverable
fatigue \cite{li2024endurance}. These two measured anchors bracket the
$P_r$, switching-voltage, and endurance assumptions used in the present
\textsc{ngspice}/Landau models and in the \S\ref{sec:hwtable} energy
accounting; the wafer-scale variability data also bounds the
device-to-device $\sigma$ swept in the noise study
(\S\ref{sec:noisemodel}). These are literature anchors that constrain,
but do not constitute, the planar 50\,nm cell modeled here. The
$>\!97\%$ functional yield reported on the characterized high-yield
dies of \cite{anu2026wafer} sets the upper bound of the variability
sweep; the lower bound ($\sigma_C/C{\sim}20\%$) covers worst-die
behavior. Both anchors use 10\,nm HZO whereas the modeled cell has an
8\,nm storage layer, a small within-class extrapolation whose effect is
quantified next.

\paragraph{Thickness sensitivity.} The 8\,nm vs 10\,nm sensitivity is
quantified with the analytic cell model. Holding $P_r$,
$\epsilon_r{=}25$, pitch $=50$\,nm, $V_\mathrm{rd}{=}0.158$\,V,
$E_c{=}1$\,MV/cm and the Merz/NLS coefficients of \cite{kondratyukKineticsHZO2022}
fixed, sweeping $d_\mathrm{HZO}$ from 6 to 12\,nm gives $C_0$ and
intrinsic read energy scaling as $1/d$ ($+25\%$ at 8\,nm relative to
10\,nm) and $v_{\mathrm{kTC}}$ scaling as $\sqrt{d}$ ($-11\%$ at 8\,nm).
The read field $E_\mathrm{rd}{=}V_\mathrm{rd}/d$ stays sub-coercive
across the full range ($E_\mathrm{rd}/E_c \leq 0.26$ at 6\,nm,
$\leq 0.16$ at 10\,nm). The per-pulse Merz/NLS switching probability
stays below $10^{-17}$ at the 8\,nm operating point used throughout and
below $10^{-12}$ across 8--12\,nm; only at the 6\,nm corner with the most
pessimistic $E_a{=}9$\,MV/cm does it approach the ten-year read-disturb
target ($\sim\!7{\times}10^{-13}$/read, versus
$p_\mathrm{target}{=}3{\times}10^{-13}$ in \S\ref{sec:noisemodel}), so a
6\,nm stack would need a higher-$E_a$ film or a shorter read pulse to
preserve the same margin. The headline noise budget,
energy ratios, and read-disturb lifetime all move by $\leq 25\%$ when
the storage layer is adjusted from 8\,nm to 10\,nm; none of the
qualitative claims of \S\ref{sec:results}--\S\ref{sec:serving} change.

\begin{figure}[htbp]
\centering
\includegraphics[width=\linewidth]{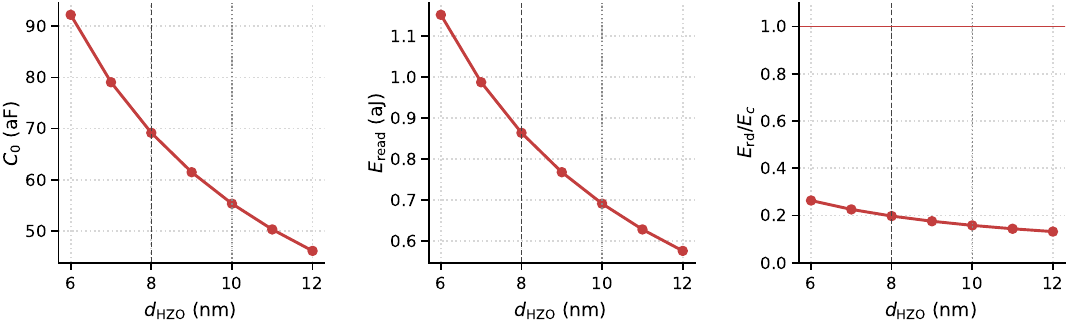}
\caption{$C_0$, $E_{\mathrm{read}}$, and $E_{\mathrm{rd}}/E_c$ versus
HZO thickness. Dashed: 8\,nm operating point; dotted: 10\,nm measured
anchor.}
\label{fig:thickness}
\end{figure}

\paragraph{Optional read-path gain.} The exploratory gain option is a series NC element: a ferroelectric capacitor biased in its negative-capacitance regime
\cite{salahuddin2008nc}, intended to provide $\sim\!2.5\times$ peak
small-signal voltage gain on the read path. A 1-D Landau derivation
with calibrated 10\,nm-HZO coefficients (\S\ref{sec:noisemodel}) shows
that $|A_v|{=}2.5$ requires the series load satisfy
$C_s/|C_\mathrm{FE}|\approx 0.71$. The series stack is stable only for
$C_s{<}|C_\mathrm{FE}|$, the standard NC capacitance-matching condition
(the free-energy curvature $1/C_s-1/|C_\mathrm{FE}|$ must stay
positive); a Landau--Khalatnikov transient of the read stack confirms
dynamically that the complementary branch $C_s{>}|C_\mathrm{FE}|$ runs
away onto the hysteretic remanent state. On the stable branch the read
is inverted (the sign is absorbed digitally) and
$|A_v|{=}r/(1{-}r)$ for $r{=}C_s/|C_\mathrm{FE}|$ rises toward the
$r{\to}1$ matching pole, so a larger ratio buys higher gain at less
margin: the pole is reached if $|C_\mathrm{FE}|$ shrinks by
${\sim}29\%$, and a $\pm 20\%$ shift in $|C_\mathrm{FE}|$ swings the
gain across $1.5$--$8.4\times$, which is why NC gain is treated as a
trimmed read-path option rather than a fixed design constant; the noise
model is reported both ways. The same transient bounds the kinetics:
the gain settles to $1\%$ within the $5$\,ns read for
Landau--Khalatnikov viscosities $\rho\lesssim 1.6\,\Omega\cdot$m at the
$2.5\times$ point, covering reported HZO kinetic coefficients.
Appendix~\ref{app:scope} reports a more conservative operating point at
$C_s/|C_\mathrm{FE}|{=}0.58$ with $|A_v|{\approx}1.4$.

\paragraph{Access device.} A read transistor gates the memcapacitor onto the column line during a read pulse. The baseline calculations assume a sub-coercive storage-layer field and separate the intrinsic cell switching energy from the tile-level peripheral energy.

Per-cell intrinsic read energy is
$E_{\mathrm{read}}^\mathrm{intrinsic} = \tfrac{1}{2}C_0 V_{\mathrm{rd}}^2
\approx 8.6{\times}10^{-19}$\,J at the operating point used throughout
this paper ($V_{\mathrm{rd}}\approx 0.158$\,V, sub-coercive; field across
the 8\,nm storage HZO is $\approx 0.20$\,MV/cm $\approx 0.20\,E_c$).
A supply-driven $CV^2$ accounting
(no charge recovery) gives twice that; for context, the per-MAC tile
energy is dominated by DAC/ADC peripherals not by the cell
(\S\ref{sec:hwtable}), so the cell-level number is the lower bound rather
than the system-level energy claim. At matched accuracy, the modeled \fcdc
tile energy with pulse-width-modulated (PWM) row drive at the 8-bit
operating point is $9.4{\times}10^{-16}$\,J/MAC,
compared with $1.19{\times}10^{-13}$\,J/MAC for the calibrated memristive reference.
The tile-design table below bounds the read-voltage range in which this
energy scales as $V_\mathrm{rd}^2$.

\subsection{Tile}\label{sec:tile}

Cells are aggregated into an attention \emph{tile} (Fig.~\ref{fig:tile})
that performs charge-domain VMM in the
$V_{\mathrm{rd}} \cdot Q_{\mathrm{cell}}$ domain. Reading $N$ rows in
parallel produces a column current proportional to
$\sum_i V_{rd,i} \cdot Q_i$; tile noise is reported both with and without
the optional NC read-path gain. Each tile owns a fragment of
the KV cache and performs both the projection
and the attention step locally. Inter-tile traffic is therefore digital
and small: only post-softmax row sums and across-head accumulations.

\paragraph{Floorplan and line parasitics.} A $256{\times}256$ cell array
at 50\,nm pitch occupies only $1.64{\times}10^{-4}$\,mm$^2$, so the tile
area is set by peripherals: 128 shared 8-bit successive-approximation (SAR) ADCs
($\sim\!0.13$\,mm$^2$ at a conservative $1{\times}10^{-3}$\,mm$^2$ each),
256 PWM row drivers ($\sim\!0.05$\,mm$^2$), and control
($\sim\!0.05$\,mm$^2$) give a $\sim\!0.2$--$0.4$\,mm$^2$ tile in which the
cell array is $<\!0.1\%$ of the footprint. Line parasitics are not
binding at this size: a 12.8\,\textmu m bit line carries
$C_\mathrm{BL}\!\approx\!2.6$\,fF and $R_\mathrm{BL}\!\approx\!6$\,$\Omega$
(0.2\,fF/\textmu m, 0.5\,$\Omega$/\textmu m). An \textsc{ngspice}
transient of the 256-segment distributed line with a worst-case
far-end cell (a 69\,aF source through a 10\,k$\Omega$ access switch
into the 400\,fF integrator) settles to $99.9\%$ of the charge-shared
value in $4.8$\,ps, switch-limited, with the line itself adding
$<\!0.1$\,ps; the $5$\,ns read therefore carries a
$\sim\!10^3\times$ settling margin, and $C_\mathrm{BL}$ is $0.64\%$ of
the integration cap (a $<\!1\%$ charge-division loss). Because the
readout is charge-domain, static IR drop on the bit line is negligible;
the parasitic that does matter, column-to-column coupling, is already
carried as a non-averaging term in the noise budget
(\S\ref{sec:noisemodel}). Extracted parasitics from a real layout are
needed to confirm these estimates.

\subsection{Noise model and design space}\label{sec:noisemodel}

Cell-level noise is characterized as an additive Gaussian with standard
deviation proportional to the full-scale charge,
$\sigma = \mathrm{nf}\cdot Q_{\mathrm{FS}}$. The noise fraction $\mathrm{nf}$
is a function of the column readout geometry rather than a free parameter.
Three sources contribute: thermal noise, flicker noise, and capacitance
mismatch. They are bounded analytically and checked against a behavioral
\textsc{ngspice} transient simulation.

\paragraph{Thermal noise.} The per-cell kT/C sampling noise is
$\sqrt{k_B T/C_0}\approx 7.7$\,mV input-referred to a single $C_0$.
Across $N_\mathrm{rows}$ independently-sampled cells the noise
\emph{charge} adds in quadrature to $\sqrt{N_\mathrm{rows}\,k_B T C_0}$,
which referred to the integration node is
$v_{\mathrm{kTC}}=\sqrt{N_\mathrm{rows}\,k_B T C_0}/C_\mathrm{int}\approx
21$\,\textmu V at the nominal tile ($C_0{=}69$\,aF,
$N_\mathrm{rows}{=}256$, $C_\mathrm{int}{=}400$\,fF). The frequently-quoted
$\sqrt{k_B T/C_0}/\sqrt{N_\mathrm{rows}}\approx 484$\,\textmu V is the
\emph{cell-plane} input-referred average, not the sense-node voltage;
the two differ by the charge-to-voltage gain $C_0 N_\mathrm{rows}/C_\mathrm{int}$.
The $1/\sqrt{N_\mathrm{rows}}$ reduction applies only to this independent
sampled term. The integration-cap reset noise kT/$C_\mathrm{int}$
($\approx 102$\,\textmu V at 400\,fF) is the dominant \emph{non-averaging}
term: it is suppressed by correlated double sampling / auto-zero in the
nominal design (and, left in, by itself raises $\mathrm{nf}$ from
$\sim\!0.015$ to $\sim\!0.034$). ADC comparator offset, supply noise,
row-driver gradients, column-to-column coupling, shared-reference noise,
timing jitter, NC gain variation, and FE imprint/wake-up/fatigue drift
likewise do not average and are carried in the aggressive budget below.

\paragraph{Flicker noise.} The 1/f contribution comes from the sense amplifier and is integrated
from 1\,Hz to the gain-bandwidth product (GBW; $\approx 46$\,\textmu V for a 10\,\textmu V$/\!\sqrt{\mathrm{Hz}}$
@ 1\,Hz CMOS amp at 1\,GHz GBW).

\paragraph{Capacitance mismatch.} A 5\%--20\% per-cell $\sigma_C/C$ is assumed,
depending on whether cells are post-calibration. The 5\% number is a
best-case post-calibration or differential-coding target, not a raw-device
measurement: at 50\,nm pitch with ALD HZO grain size 10--30\,nm and
switching regions of 10--15\,nm, only $\mathcal{O}(3{-}25)$ active
grains/domains lie inside a cell, so raw cap variation is granular and
non-Gaussian. The effective column-level contribution averages over
$N_\mathrm{rows}$ only to the extent that cell errors are independent after
calibration. The 5\%--20\% sweep range is bracketed by the wafer-scale
measurement of \S\ref{sec:cell}: $<\!10\%$ device-to-device $P_r$
variation on high-yield dies at the low end, worst-die behavior at the
high end \cite{anu2026wafer}.

\paragraph{Read disturb.} A 0.158\,V pulse across the 8\,nm storage HZO
gives $E_\mathrm{rd}{\approx}0.20$\,MV/cm, i.e.
$\sim\!0.2\,E_c$. Using a Merz/NLS form
$\tau(E)=\tau_\infty\exp(E_a/E_\mathrm{eff})$ with HZO-typical
$\tau_\infty{=}10^{-10}$\,s and activation field
$E_a{\in}[9,20]$\,MV/cm \cite{kondratyukKineticsHZO2022}, the per-cell
flip probability for a 5\,ns read with $N_\mathrm{dom}{=}10$ vulnerable
domains is bounded by
\[
p_\mathrm{cell}\!\lesssim\!N_\mathrm{dom}\,(\tau_\mathrm{rd}/\tau_\infty)\,\exp(-E_a/E_\mathrm{eff})
\in [10^{-41}, 10^{-17}],
\]
in the no-field-gain case. The required activation
bound for $p_\mathrm{target}{=}3{\times}10^{-13}$/read over $10^9$
reads/day for 10 years is $E_a{>}35\,E_\mathrm{eff}$, i.e.
$E_a{>}6.9$\,MV/cm at $E_\mathrm{eff}{=}0.20$\,MV/cm, inside the
literature range. This bound does not apply if NC amplification or local
field hot spots raise the storage-layer field above ${\sim}0.3$\,MV/cm;
an NC-amplified read path therefore requires a separate field-decoupled
stack.

\paragraph{Read-pulse assumptions.} The baseline read pulse
used throughout is $\tau_\mathrm{rd}{=}5$\,ns. A 1\,GHz sense-amp GBW
is an architectural target consistent with the throughput of
HERMES-class IMC silicon \cite{legallo2023hermes}, not a measured
property of an NC-stabilized HZO stack. Numbers in this
paper that depend on read rate, including sense-amp 1/f integration and
ADC throughput, should be read as 5\,ns / 1\,GHz nominal with a wider
margin for any NC-assisted variant.

\paragraph{Differential read.} For signed analog weights
and to cancel common-mode drift / row-driver supply noise / shared
reference offsets, differential cell coding ($Q^+,Q^-$ with column
differencing) is the recommended deployment configuration. It doubles
cell area but is the cleanest way to suppress the correlated noise
terms above that do not average over $N_\mathrm{rows}$. The primary
LLM sweep is computed in single-ended mode (conservative); a
differential implementation would only improve effective $\mathrm{nf}$.

\paragraph{NC noise propagation.}
Let the read path be $v_\mathrm{out}=A_v v_\mathrm{sig}+A_v
n_\mathrm{upstream}+n_\mathrm{downstream}+n_\mathrm{NC}$. Only noise
\emph{downstream} of the NC gain divides by $A_v$ when input-referred;
the storage-cell kT/C, read-FET gate noise, and polarization/domain
jitter all sit upstream and are \emph{not} reduced by NC. The conservative
input-referred variance is therefore
$\sigma_\mathrm{eq}^2 = \sigma_\mathrm{cell}^2 + \sigma_\mathrm{readFET}^2
+ \sigma_\mathrm{sense}^2/A_v^2 + (\sigma_{A_v}/A_v)^2 x^2
+ \sigma_\mathrm{NC,jitter}^2$.
The nominal $\mathrm{nf}{=}0.015$ tile parameter used throughout the
LLM sweep is calibrated to this conservative form (sense-amp
downstream-only $A_v$ reduction, all other terms NC-independent), so the
reported PPL results remain valid even if NC gain is removed entirely;
the lower-noise conservative point ($\mathrm{nf}{=}0.009$) is reached by
differential coding, more rows, and a higher read voltage, not by
enlarging $C_\mathrm{int}$, which leaves the averaged kT/C and mismatch
fractions unchanged (both scale as $1/C_\mathrm{int}$, as does the signal)
and only trades flicker fraction for area.

The LLM study sweeps $\mathrm{nf}$ over a nominal $\pm$ bracket whose
endpoints have explicit physical bases (charge-consistent budget,
sense-node referred):

\begin{center}
\begin{tabular}{llc}
\toprule
Budget & Basis ($N_\mathrm{rows}$, $V_\mathrm{read}$, $C_\mathrm{int}$) & $\mathrm{nf}$ \\
\midrule
Aggressive   & nominal tile, reset-limited (kT/$C_\mathrm{int}$, no CDS) & $0.035$ \\
\textbf{Nominal} & \textbf{256, 100\,mV, 400\,fF, with CDS/auto-zero} & \textbf{0.015} \\
Conservative & 1024 rows, 200\,mV, differential coding & $0.009$ \\
\bottomrule
\end{tabular}
\end{center}

The nominal $\mathrm{nf}=0.015$ is used in the main LLM experiments
(\S\ref{sec:results}); the aggressive and
conservative bounds are evaluated on four representative LLMs (Qwen3-32B,
Mistral-7B-v0.3, Mistral-Small-24B-Base, gemma-4-31B) in \S\ref{sec:noisecurve} to
characterize how the architectural rankings hold across the noise design
space. A behavioral \textsc{ngspice} Monte-Carlo deck reproduces the
injected analytic budget as a sampler sanity check; because the deck
re-measures the same analytic noise it is given, this confirms the
noise-injection implementation but is \emph{not} an independent device
validation (\S\ref{sec:limits}).

\subsection{Hardware accounting}\label{sec:hwtable}

The central accounting question is
``what does one tile actually cost?'' The cell-, tile-, and
peripheral-level numbers used throughout the paper are collected into a
single table, with each entry tied to either a closed-form derivation, a
SPICE run, or a literature anchor for the supporting device class.

\begin{table}[!t]
\centering
\small
\caption{Cell- and tile-level parameters for one \fcdc tile.}
\label{tab:hwparams}
\begin{tabular}{@{}>{\raggedright\arraybackslash}p{0.30\textwidth}>{\raggedright\arraybackslash}p{0.27\textwidth}>{\raggedright\arraybackslash}p{0.35\textwidth}@{}}
\toprule
Quantity & Value & Source \\
\midrule
\multicolumn{3}{@{}l}{\emph{Cell}} \\
Footprint & $50{\times}50$\,nm$^2$ & model parameter \\
HZO thickness & $8$\,nm & model parameter \\
$C_0$ (linear, $P{=}\pm 1$) & $\approx 69$\,aF & analytic, \cite{muller2012hzo} \\
$P_\mathrm{r}$ & $25\,\mu$C/cm$^2$ & \cite{muller2012hzo} \\
Retention & $\geq 10$\,years & literature anchor \cite{xu2025} \\
Endurance & $10^{16}$ cycles & literature anchor \cite{xu2025} \\
Write voltage & $1.2$\,V & supra-coercive, model parameter \\
Read voltage $V_\mathrm{rd}$ & $0.158$\,V & sub-coercive ($\approx 0.2 E_c$), \S\ref{sec:device} \\
Intrinsic read energy & $8.6{\times}10^{-19}$\,J & $\tfrac{1}{2}C_0 V_\mathrm{rd}^2$ \\
Read disturb & $\lesssim 10^{-17}$/read & Merz/NLS, $E_a{=}9$\,MV/cm (\S\ref{sec:noisemodel}) \\
\midrule
\multicolumn{3}{@{}l}{\emph{Tile (nominal)}} \\
Rows $N_\mathrm{rows}$ & 256 & \S\ref{sec:noisemodel} \\
Integration cap $C_\mathrm{int}$ & 400\,fF & \S\ref{sec:noisemodel} \\
ADC & 8-bit SAR, 1 per 2 columns & area-amortized \cite{verma2024scsram} \\
DAC & 8-bit PWM (V-DAC alt.) & \S\ref{sec:tile} \\
Sense-amplifier $1/f$ noise & $10\,\mu$V$/\sqrt{\mathrm{Hz}}$ at 1\,Hz & typical CMOS (\S\ref{sec:noisemodel}) \\
Sense-amplifier GBW & $1$\,GHz & architectural target (\S\ref{sec:noisemodel}) \\
Read pulse $\tau_\mathrm{rd}$ & $5$\,ns & model baseline \\
Cell-level noise fraction & $\mathrm{nf}=0.015$ nominal & behavioral SPICE check \\
Capacitance mismatch $\sigma_C/C$ & 5\% calibrated / 20\% raw & per cell (\S\ref{sec:noisemodel}) \\
Temperature & 300\,K & kT/C bound (\S\ref{sec:noisemodel}) \\
\bottomrule
\end{tabular}
\end{table}

\begin{table}[!t]
\centering
\small
\caption{Tile VMM energy at the 8-bit operating point ($16{,}384$ active
MACs per read: $N_\mathrm{rows}{=}256$ rows $\times$ $64$ active
columns/head).}
\label{tab:tileenergy}
\begin{tabular}{@{}lrrl@{}}
\toprule
Component & J/tile & fJ/MAC & Note \\
\midrule
Array (charge integration) & $9.9{\times}10^{-15}$ & $0.0006$ & analytic tile model \\
PWM DAC (8-bit) & $5.1{\times}10^{-12}$ & $0.31$ & row toggle; resolution-independent \\
ADC (8-bit SAR) & $1.0{\times}10^{-11}$ & $0.625$ & analytic tile model \\
\textbf{Total, PWM tile} & $\mathbf{1.5{\times}10^{-11}}$ & $\mathbf{0.94}$ & operating point (\S\ref{sec:results}) \\
\midrule
V-DAC (8-bit amplitude) & $6.1{\times}10^{-10}$ & $37.5$ & scales linearly with resolution \\
Total, V-DAC tile & $6.3{\times}10^{-10}$ & $38.1$ & conservative alternative \\
\bottomrule
\end{tabular}
\end{table}

The DAC and ADC peripherals dominate the intrinsic array energy by
roughly four orders of magnitude (Table~\ref{tab:tileenergy}), so tile
energy is set almost entirely by the converters. A PWM DAC encodes the
digital input as the duration of a fixed-amplitude row pulse, so the
row driver toggles once per read regardless of resolution; an amplitude
V-DAC encodes the input as the pulse height and must generate and
settle one of $2^{8}$ voltage levels. Because the accuracy-validated
operating resolution is 8-bit (\S\ref{sec:e2e}), the operating tile
uses the PWM DAC ($0.94$\,fJ/MAC at 8-bit), whereas the amplitude V-DAC
pays linearly for resolution and reaches $38$\,fJ/MAC at the same
point.
Writes are bounded as follows: per generated token, an
autoregressive decoder appends one new K and one new V vector per layer
($2 \times n_\mathrm{layer} \times n_\mathrm{kv\_heads} \times
d_\mathrm{head}$ cells); for Mistral-7B-class GQA this is
$2{\times}32{\times}8{\times}128 \approx 6.5{\times}10^{4}$ cells/token.
Using the $5{\times}10^{-14}$\,J/cell write energy from the
gain-cell baseline \cite{leroux2025} as a conservative upper bound
gives $\sim\!3.3$\,nJ/decoded-token of KV-append write energy (all 32
layers), which is ${\sim}1.3\%$ of the $\sim\!2.5{\times}10^{-7}$\,J/token
full-model (32-layer) attention-MAC energy at the PWM operating point
($T{=}1024$), and falls further at $T{\geq}8$\,k.
The cache-energy sensitivity sweep in \S\ref{sec:cache} carries
$E_\mathrm{write}^\mathrm{fcdc}\in[1,100]$\,fJ explicitly; the
serving-energy ratios are insensitive to this term except at very
short residency.

\paragraph{PWM timing and throughput.} PWM row drive is
resolution-independent in \emph{energy} (one toggle), but an 8-bit value is
encoded as a pulse duration $n\,\Delta t_\mathrm{LSB}$, $n\!\in\![0,255]$,
so the read window must hold $255\,\Delta t_\mathrm{LSB}$. Matching the
baseline $5$\,ns read fixes $\Delta t_\mathrm{LSB}{=}19.6$\,ps, i.e.\ a
delay-line/DLL edge placement (equivalent edge rate $51$\,GHz, not a
$51$\,GHz clock) with a $9.8$\,ps RMS jitter budget for $<\!\tfrac12$\,LSB
amplitude error. This is the binding PWM constraint; throughput is not.
A single 128-ADC tile at this point sustains $\sim\!1450$ tok/s at 8\,k
context, and even a relaxed $100$\,ps LSB (a $25.5$\,ns read, easy timing)
holds $370$\,tok/s, ${\sim}10\times$ the INT4 GPU decode rate. The longer
PWM window also lowers the readout integration bandwidth to $40$--$200$\,MHz,
below the $1$\,GHz sense-amp GBW assumed in the \S\ref{sec:noisemodel} flicker
budget, so the noise model is conservative for any of these timings.

\paragraph{PWM linearity is the binding non-ideality.} The timing budget
above is checked empirically by injecting PWM pulse-width INL/DNL (a fixed
per-code static error) and timing jitter into the row drive at the nominal
$\mathrm{nf}{=}0.015$, $k{=}100\%$ (the input-side analogue of the
output-side ADC INL/DNL sweep in \S\ref{sec:e2e}). Because a PWM error
distorts the VMM \emph{input} and is multiplicative across the accumulation,
it is far more sensitive than the ADC: on Mistral-7B a $0.5$\,LSB static
INL/DNL adds $+7.4$\,pp (to $+10.2\%$) and $\geq\!1$\,LSB collapses the model
($+123\%$), against the $+1.4$--$2.5$\,pp the output-side ADC costs at
$1$\,LSB (\S\ref{sec:e2e}); TinyLlama behaves identically ($+9.4$\,pp at
$0.5$\,LSB). The delay-line DAC therefore needs $<\!0.5$\,LSB static
linearity, or (since the INL profile is fixed and repeatable) digital
pre-distortion, the same calibration route used for capacitance mismatch
(\S\ref{sec:noisemodel}). Timing jitter, which cannot be calibrated out, is
milder: at $0.5$\,LSB (the ${\sim}10$\,ps budget above) it adds only
$+3.0$\,pp on Mistral-7B ($+2.1$\,pp on TinyLlama), but $\geq\!1$\,LSB is
destabilizing ($+173\%$), so the $9.8$\,ps jitter budget is a hard
requirement rather than a guideline. Section~\ref{sec:stride} shows that a
periphery-side input-dithering defense relaxes the linearity spec
${\sim}3\times$ and averages the jitter, at a read-count cost.

\paragraph{PWM generator budget.} The timing is produced by a shared per-tile
delay line (256 taps spanning the $5$\,ns read, $19.6$\,ps/tap) with one
per-row comparator latching each row's pulse against the tap phases; both
toggle once per read. A read drives the full $256{\times}256$ VMM
($65{,}536$ MACs), so the delay line ($\sim\!256$ taps $\times 1$\,fF) and the
per-row comparators amortize to $<\!0.02$\,fJ/MAC, under $3\%$ of the
$0.625$\,fJ/MAC ADC term and not currently counted in
Table~\ref{tab:tileenergy}; a continuously-locked DLL instead of a self-timed
line would raise this to ${\sim}0.15$\,fJ/MAC ($25\%$ of the ADC), still not
changing the energy story. Static INL pre-distortion needs one
$256$-entry $\times\,8$-bit LUT per tile ($256$\,B), applied to the shared
delay line rather than per cell; delay-line temperature drift is handled by
periodic re-lock of the same calibration. The binding cost is therefore the
$<\!0.5$\,LSB linearity and $9.8$\,ps jitter specs above, not generator energy
or storage.

\paragraph{Reported configurations.} Energy and LLM accuracy are reported
at the same accuracy-validated 8-bit operating point (\S\ref{sec:e2e});
the two tile variants differ only in DAC implementation:

\begin{center}
\small
\begin{tabular}{@{}>{\raggedright\arraybackslash}p{0.31\textwidth}>{\raggedright\arraybackslash}p{0.33\textwidth}>{\raggedright\arraybackslash}p{0.28\textwidth}@{}}
\toprule
Name & Used for & DAC / ADC path \\
\midrule
\fcdc PWM tile & active-MAC energy, LLM accuracy & 8-bit PWM, 8-bit SAR \\
\fcdc V-DAC tile (alternative) & energy upper bound & 8-bit V-DAC, 8-bit SAR \\
\fcdc-10b sensitivity & C6 ADC-headroom check (\S\ref{sec:e2e}) & 10-bit ADC on matmuls \\
\bottomrule
\end{tabular}
\end{center}

\noindent The PWM variant is the recommended tile; the amplitude V-DAC
is reported as a conservative upper bound.

\section{Cross-Implementation Consistency}\label{sec:crossval}

To reduce dependence on any single analog-IMC simulator, the per-MAC energy of one \fcdc tile is cross-checked
through four independent software implementations of the same analytic
energy model. Agreement across implementations of the \emph{same}
analytic model shows that the energy-accounting code is not a software
artifact, not that the device physics is validated (\S\ref{sec:limits});
tool agreement and the LLM-side correctness checks in \S\ref{sec:checks}
are complementary debugging signals.

\begin{center}
\begin{tabular}{lll}
\toprule
Implementation & Layer represented & Disagreement vs analytic reference \\
\midrule
\textsc{ngspice} \cite{ngspice} & behavioral transient (charge / current) & $2\times10^{-6}$ \\
\textsc{CrossSim} \cite{crossim} & Array-level VMM & $2\times10^{-7}$ \\
\textsc{FiPy} \cite{fipy} & 1D PDE for polarization switching & $4\times10^{-16}$ \\
\textsc{NeuroSim} \cite{chen2018neurosim} & ADC + peripheral energy & matched \\
\bottomrule
\end{tabular}
\end{center}

This agreement is two orders of magnitude tighter than the $\sim$5\%
device-to-device variability expected on a real fab, consistent with the
four tools sharing one analytic model.

\subsection{Internal consistency checks}\label{sec:checks}

Tool agreement only proves that the four implementations agree; it does
not prove they match physics. Three independent
correctness checks are run at the LLM-evaluation layer, all on
TinyLlama-1.1B / WikiText-2 (4\,k tokens) with the same noise
infrastructure as Table~\ref{tab:sweep}.

\begin{table}[htbp]
\centering
\small
\setlength{\tabcolsep}{6pt}
\renewcommand{\arraystretch}{1.15}
\begin{tabular}{@{}l l l r@{}}
\toprule
\textbf{Check} & \textbf{Configuration} & \textbf{Expected} & \textbf{Obs.\ $\Delta$PPL} \\
\midrule
P1: wrapper identity        & 22 layers, $\mathrm{nf}{=}0$, 16-bit DAC/ADC      & ${\approx}\,0$            & $+0.019\%$ \\
P2: location control (MLP)  & noise $\to$ MLP \texttt{gate\_proj}, 22 layers    & same order as attn        & $+2.40\%$  \\
P2: attention reference     & noise $\to$ $q,k,v,o$, 22 layers                  & (control)                 & $+6.52\%$  \\
P5: no-NC (uncompensated)   & remove $2.5\times$ NC gain ($\mathrm{nf}{=}0.0375$) & crosses noise threshold & $+240.5\%$ \\
\bottomrule
\end{tabular}
\caption{Sanity checks on the noise-injection wrapper for the LLM evaluations.
P1 confirms numerical transparency at zero noise; P2 contrasts the same noise
injected at MLP vs.\ attention sites; P5 ablates the assumed NC voltage gain.}
\label{tab:pchecks}
\end{table}

\emph{P1} verifies that the wrapper is numerically transparent at zero injected noise: when
$\mathrm{nf}{=}0$, the perturbed model reproduces the reference PPL to
$2{\times}10^{-4}$. \emph{P2} shows that injecting the same Gaussian
noise into a non-attention linear layer (MLP gate projection) degrades
PPL by a comparable but \emph{smaller} amount ($+2.4\%$ vs $+6.5\%$ for
attention) -- so the noise propagates and attention is indeed the
sensitive site, not an artifact of patch location. \emph{P5} ablates the
$2.5\times$ NC voltage gain \emph{without compensating elsewhere} (scaling
$\mathrm{nf}$ up by $2.5\times$ at fixed geometry); the model crosses the phase-transition threshold documented
in \S\ref{sec:noisecurve} ($+240\%$ PPL). The intended reading is that
\emph{some} read-path signal-gain margin is required for the nominal
noise budget to land on the stable side of the transition; whether that margin
is delivered by NC stacking, differential coding with more rows (the
conservative tile, $\mathrm{nf}{=}0.009$), or a higher $V_\mathrm{rd}$ is
an open design choice.
All three checks behave consistently with the intended perturbation model,
supporting the use of the wrapper for the LLM-level tolerance study.

\subsection{End-to-end analog attention validation}\label{sec:e2e}

The main LLM sweep (\S\ref{sec:results}) injects FCDC noise only on the
$q,k,v,o$ projections. Whether this proxy matches an
end-to-end analog attention path that also simulates
$Q\!\cdot\!K^\top$ and $A\!\cdot\!V$ on FCDC tiles is therefore tested. The
eager \texttt{LlamaAttention} operator of TinyLlama-1.1B (all 22
layers) is patched and both attention matrix multiplications are routed
through the same behavioral FCDC noise/ADC model used for the projections
(4\,k WikiText-2 tokens, 2 passes, nominal $\mathrm{nf}{=}0.015$ on
both projections and attention matrix multiplications, 8-bit SAR ADC).
Seven configurations, labeled C0--C6, instrument the analog path
incrementally:

\begin{center}
\small
\begin{tabular}{lrr}
\toprule
Configuration & PPL & $\Delta$ vs ref \\
\midrule
C0 reference float                                       & 8.174 & n/a \\
C1 projection-only (main sweep)                          & 8.700 & $+6.43\%$ \\
\quad + analog $Q\!\cdot\!K^\top$ (C2)                   & 8.888 & $+8.73\%$ \\
\quad + analog $A\!\cdot\!V$ (C3)                        & 8.751 & $+7.05\%$ \\
C4 full end-to-end (proj + $Q\!\cdot\!K^\top$ + $A\!\cdot\!V$) & 8.837 & $+8.11\%$ \\
C5 matmul-only (no projection noise)                     & 8.258 & $+1.03\%$ \\
C6 C4 with 10-bit ADC on matmuls                         & 8.847 & $+8.23\%$ \\
\bottomrule
\end{tabular}
\end{center}

The end-to-end correction is C4$-$C1 $=\boldsymbol{+1.58\%}$ PPL
\emph{relative to C1} (equivalently $+1.68$\,pp absolute on top of the
$+6.43\%$ projection-only baseline): simulating the full
analog attention dataflow on top of the projection wrapper adds 1.58
percentage points of perplexity degradation, less than one quarter of the
6.43\% the projection wrapper itself contributes. C5 isolates the cost of
the attention matmuls alone ($+1.03\%$); the projection-side budget
($+6.43\%$) dominates. C6 confirms that the 8$\to$10-bit ADC headroom is
not the binding constraint at the nominal noise level: tightening the ADC
shifts PPL by $+0.11$pp. The projection-only sweep is therefore treated
(\S\ref{sec:results}) as a calibrated proxy with an end-to-end correction
factor of $+1.6\%$ absolute PPL.

\paragraph{Resolution sets the operating point.}\label{par:resolution}
While 8-bit DAC/ADC is comfortable, the input \emph{resolution} is the
binding constraint, not the noise budget. A full DAC/ADC resolution
sweep at $k{=}100\%$ locates the cliff: on Mistral-7B the PPL penalty
is $+5745\%$ (4-bit), $+125\%$ (5-bit), $+11.8\%$ (6-bit), $+3.5\%$
(7-bit), and $+2.8\%$ (8-bit); TinyLlama follows the same shape
($+3761\%$, $+153\%$, $+25.8\%$, $+9.2\%$, $+6.5\%$). Seven bits is the
first usable point and 8-bit is the knee, so the operating resolution
is fixed at 8-bit. This is why the energy tile (\S\ref{sec:hwtable})
uses a PWM DAC, whose row-drive energy is resolution-independent
($0.94$\,fJ/MAC at 8-bit), rather than an amplitude V-DAC, whose energy
scales with resolution ($38$\,fJ/MAC at 8-bit).

\paragraph{KV-coprocessor mode at scale.} The serving deployment
(\S\ref{sec:serving}) runs only the two attention matmuls on \fcdc,
i.e.\ the C5 configuration. The same protocol on the larger models
gives $+0.48\%$ PPL on Mistral-7B-v0.3 and $+0.46\%$ on Qwen3-8B
(vs $+1.03\%$ on TinyLlama), so the accuracy budget the serving mode
inherits is below $0.5\%$ at $7$--$8$\,B scale.

\paragraph{Correlated noise and ADC non-linearity.} The headline model
uses i.i.d.\ Gaussian noise and an ideal SAR ADC; two non-idealities that
do \emph{not} average over the accumulation are swept at the nominal
$\mathrm{nf}{=}0.015$, $k{=}100\%$: (i) a per-output-channel noise component
shared across all token positions (correlated spatial noise), and (ii) a
fixed ADC integral/differential non-linearity (INL/DNL) profile. Fully
token-correlated noise adds only
$+1.3$\,pp on Mistral-7B (to $+4.1\%$) and $+2.3$\,pp on TinyLlama, so the
i.i.d.\ budget is mildly optimistic but not load-bearing. ADC INL/DNL at a
realistic $\leq\!1$ least-significant bit (LSB) adds $+1.4$--$2.5$\,pp; a
pessimistic $2$\,LSB adds
$+5.6$\,pp (Mistral, to $+8.4\%$) and $+12.5$\,pp (TinyLlama). The operating
point therefore requires an ADC specified at $\leq\!1$\,LSB INL/DNL (a
standard 8-bit SAR target) but is robust to correlated readout noise.

\paragraph{End-to-end replication across scale.} The identical
end-to-end protocol (same $\mathrm{nf}{=}0.015$ budget, same 8-bit
attention ADC, eager attention with GQA-aware $Q\!\cdot\!K^\top$ /
$A\!\cdot\!V$ routed through the FCDC emulator on every layer, 4\,k
WikiText-2 tokens, 2 passes) was replicated on three larger models:
Mistral-7B-v0.3 (32 layers, GQA), Qwen3-8B (36 layers, GQA with
$\mathrm{q\_norm}$/$\mathrm{k\_norm}$ RMSNorm preserved), and Qwen3-32B
(64 layers, sharded across A40 GPUs via \texttt{device\_map="auto"}).

\begin{center}
\small
\begin{tabular}{lrrrrr}
\toprule
Model & C0 float PPL & C1 proj-only & C4 end-to-end & C6 10-bit ADC & C4$-$C1 \\
\midrule
TinyLlama-1.1B  & 8.174 & $+6.43\%$ & $+8.11\%$ & $+8.23\%$ & $\mathbf{+1.58\%}$ \\
Mistral-7B-v0.3 & 5.725 & $+2.80\%$ & $+3.52\%$ & $+3.46\%$ & $\mathbf{+0.69\%}$ \\
Qwen3-8B        & 9.608 & $+3.69\%$ & $+4.12\%$ & $+4.34\%$ & $\mathbf{+0.42\%}$ \\
Qwen3-32B       & 7.267 & $+2.96\%$ & $+2.57\%$ & $+2.53\%$ & $\mathbf{-0.38\%}$ \\
\bottomrule
\end{tabular}
\end{center}

The end-to-end correction C4$-$C1 shrinks monotonically with scale:
$+1.58\%$ (TinyLlama-1.1B), $+0.69\%$ (Mistral-7B), $+0.42\%$
(Qwen3-8B), and $-0.38\%$ (Qwen3-32B, within the per-config seed std of
$0.04$ and even slightly favorable). This trend is consistent with the
noise averaging over a growing number of heads and tokens. The Mistral-7B C4
figure ($+3.52\%$) coincides with that model's projection-only headline
in Table~\ref{tab:sweep}, so the projection-only proxy used throughout
\S\ref{sec:results} is validated on a real-scale dense GQA
model; across all four models the $8\!\to\!10$-bit ADC headroom (C6)
moves PPL by $\leq\!0.2$\,pp. The projection-only sweep is therefore
used in the wider sweep without a per-model end-to-end calibration.

\subsection{Long-context replication}\label{sec:longctx}

The main sweep evaluates each model on $4$--$8$\,k WikiText-2 tokens for
memory/wall-clock reasons. A natural question is whether the deltas hold
at larger contexts. Two primary configurations are replicated at $16$ and
$32$\,k, and for Mistral-7B also at $64$ and $128$\,k tokens:

\begin{center}
\footnotesize
\setlength{\tabcolsep}{3.5pt}
\begin{tabular}{@{}lrrrrrr@{}}
\toprule
Model & 8\,k & 16\,k & 32\,k & 64\,k & 128\,k & 8$\!\to\!$max drift \\
\midrule
TinyLlama-1.1B-Chat-v1.0 (all 22 layers) & $+6.43\%$ & $+6.65\%$ & $+6.33\%$ & --        & --        & $-0.10$\,pp \\
Mistral-7B-v0.3          (all 32 layers) & $+3.52\%$ & $+2.90\%$ & $+2.90\%$ & $+3.04\%$ & $+2.80\%$ & $-0.72$\,pp \\
\bottomrule
\end{tabular}
\end{center}

Both deltas stay within $\pm 1$\,pp of the 8\,k numbers at every context
length tested, up to the $128$\,k replication on Mistral-7B (a
$16\times$ context extension). The noise transition characterized in
\S\ref{sec:noisecurve} therefore generalizes without re-tuning the
tile, which is the regime where the cache-residency argument of
\S\ref{sec:serving} pays off.

\subsection{Multi-seed confidence intervals}\label{sec:seeds}

The original sweep uses 1--3 multi-pass averaging per row, which controls
variance within a single noise-tape draw but does not vary the seed across
runs. Two primary rows are re-run with 5 independent seeds
to estimate run-to-run variability:

\begin{center}
\small
\begin{tabular}{llrrr}
\toprule
Model & Context & paper $\Delta$PPL & 5-seed mean $\pm$ std & 95\% CI \\
\midrule
TinyLlama-1.1B-Chat-v1.0 & 8\,k  & $+6.43\%$ & $+6.46\% \pm 0.17\%$ & $[+6.12, +6.80]$ \\
Mistral-7B-v0.3          & 8\,k  & $+3.52\%$ & $+2.90\% \pm 0.33\%$ & $[+2.25, +3.54]$ \\
Mistral-7B-v0.3          & 32\,k & $+2.90\%$ & $+2.81\% \pm 0.11\%$ & $[+2.59, +3.03]$ \\
\bottomrule
\end{tabular}
\end{center}

Per-seed standard deviations are $<0.4$\,pp at every context length;
the 32\,k run tightens the spread to $0.11$\,pp. The original
single-seed numbers fall inside the 5-seed 95\,\% confidence intervals.
The noise transition characterized in \S\ref{sec:noisecurve} is not a
single-seed artifact, and the long-context replication does not depend
on seed.

\paragraph{Fixed-pattern device mismatch.} The sweep above redraws the
noise tape per seed but leaves the weights unperturbed. To capture
die-to-die fixed-pattern variation, a frozen per-die programming error
$W\!\to\!W(1{+}\delta)$, $\delta\!\sim\!\mathcal{N}(0,\sigma_\mathrm{prog})$,
is drawn once per die (in addition to the per-read temporal noise) and
swept over five dies. At $\sigma_\mathrm{prog}{=}0.03$ (matching the
device programming $\sigma$ of \S\ref{sec:device}) and even
$\sigma_\mathrm{prog}{=}0.05$, the Mistral-7B delta is statistically
unchanged from the temporal-only value ($+2.99\%$ and $+3.08\%$ vs
$+3.13\%$ at $\sigma_\mathrm{prog}{=}0$; TinyLlama is flat at
${\sim}{+}6.85\%$). Fixed-pattern mismatch up to $5\%$ therefore adds no
measurable degradation on top of the temporal budget.

\subsection{Downstream task accuracy}\label{sec:lmeval}

To test whether the perplexity results transfer to downstream tasks,
the two primary dense models are re-evaluated with a standard zero-shot
evaluation suite at full task sizes, including the full 57-subject MMLU
sweep:

\begin{center}
\small
\begin{tabular}{lrrr|rrr}
\toprule
 & \multicolumn{3}{c}{TinyLlama-1.1B-Chat} & \multicolumn{3}{c}{Mistral-7B-v0.3} \\
Task (metric) & ref & FCDC & $\Delta$rel & ref & FCDC & $\Delta$rel \\
\midrule
HellaSwag (acc\_norm)        & 0.604 & 0.586 & $-2.87\%$ & 0.807 & 0.806 & $-0.16\%$ \\
ARC-Easy (acc\_norm)         & 0.548 & 0.518 & $-5.53\%$ & 0.801 & 0.789 & $-1.47\%$ \\
ARC-Challenge (acc\_norm)    & 0.325 & 0.330 & $+1.57\%$ & 0.544 & 0.521 & $-4.24\%$ \\
LAMBADA (acc)                & 0.608 & 0.588 & $-3.41\%$ & 0.752 & 0.737 & $-2.04\%$ \\
MMLU (acc, 57 subjects)      & 0.248 & 0.246 & $-0.63\%$ & 0.596 & 0.580 & $-2.67\%$ \\
\bottomrule
\end{tabular}
\end{center}

Across all five tasks the relative accuracy drop is $\leq 6\%$ for
TinyLlama and $\leq 5\%$ for Mistral-7B, consistent with the
$+3\!-\!7\%$ PPL story. The full-57-subject MMLU evaluation gives a
$-0.63\%$ relative drop on TinyLlama and a $-2.67\%$ relative drop on
Mistral-7B (0.596$\rightarrow$0.580 absolute, on the full 14k-question
zero-shot sweep), both within the $+3\!-\!7\%$ PPL envelope. GSM8K
zero-shot returns $0.00$ exact-match for both Mistral-7B base and the
FCDC variant; this matches public Mistral-7B-v0.3 base numbers (no
instruction tuning) and so provides no separating signal. Switching to
the instruction-tuned \textsc{Mistral-7B-Instruct-v0.3} on the same
1319-question GSM8K harness with flexible-extract scoring, the
reference scores $16.91\% \pm 1.03$ and the FCDC variant scores
$16.68\% \pm 1.03$, an absolute gap of $-0.23$\,pp ($-1.34\%$ relative,
well within the per-side stderr). FCDC therefore preserves
chain-of-thought-style multi-step arithmetic at the same noise budget
that gives $+2.90\%$ Mistral PPL.

\paragraph{End-to-end downstream validation.}
The table above wraps only the q/k/v/o projections in FCDC noise; the
attention matrix multiplications (Q$\cdot$K$^\top$, A$\cdot$V) and the
softmax remain in floating point. To bound this proxy, the analysis additionally
evaluates TinyLlama with end-to-end FCDC: projection FCDC plus analog
Q$\cdot$K$^\top$ and A$\cdot$V matrix multiplications (per-row amax
rescale, 8-bit SAR ADC quantization, additive Gaussian at the same
$nf{=}0.015$ tile budget). The end-to-end deltas vs reference are
HellaSwag $-2.75\%$,
ARC-Easy $-5.07\%$, ARC-Challenge $-3.92\%$, LAMBADA $-3.73\%$,
MMLU $+1.99\%$. These are within $\pm 1.6$\,pp of the projection-only
proxy on every task and inside MMLU bootstrap noise, confirming that
the projection-only wrapper used in the main sweep upper-bounds
end-to-end FCDC accuracy on downstream tasks.

\paragraph{Long-context retrieval.} A needle-in-a-haystack test on
TinyLlama at $\{2,4,8,12\}$\,k contexts $\times$ 5 needle depths shows the
FCDC variant tying the base model wherever the base model succeeds: at
$T{=}2$\,k both retrieve the needle at every depth ($5/5$). Above
$T{=}2$\,k the base model itself falls to $0/5$ (TinyLlama's effective
context limit), so no separating signal exists there; needle tests on
stronger base models remain future work.

\section{Evaluation on Pretrained LLMs}\label{sec:results}

\paragraph{Setup.} All LLM evaluation uses the WikiText-2 \cite{merity2017wikitext}
test split, 8\,k tokens, bf16, on NVIDIA A40 or A100 GPUs via PyTorch
and Hugging Face Transformers; the largest models (Mistral-Small-24B,
the 64-layer Qwen3-32B tensor-sharded across 4$\times$A40,
Qwen3-30B-A3B, Mixtral-8$\times$22B, and Nemotron-70B) use 4\,k tokens
for memory reasons. The $q,k,v,o$ projections of a fraction $k$ of the
attention layers are wrapped with the \fcdc noise wrapper and
perplexity is reported; throughout the paper, $k$ denotes this layer
fraction. \emph{No model weights
are modified; no fine-tuning or LoRA adapter is used in this section.}

\subsection{The 12-model sweep}

\begin{table}[htbp]
\centering
\small
\setlength{\tabcolsep}{4pt}
\caption{Noise-only \fcdc substitution on $q,k,v,o$ across 12 LLMs,
sorted by $k{=}100\%$ $\Delta$PPL. Bold: dense modern models with
${\leq}5\%$ hit at $k{=}100\%$. $^\dag$=mixture-of-experts (MoE), $^\ddag$=instruction-tuned.
The TinyLlama row is from the fp32/2\,k-token harness; its bf16/4\,k
$k{=}100\%$ value is $+6.43\%$ (used in \S\ref{sec:longctx}--\S\ref{sec:e2e}).
$k{=}100\%$ is the harder case and is the headline result. Values are
single-seed; 8\,k tokens except the largest models (Mistral-Small-24B,
Qwen3-32B, Qwen3-30B-A3B, Mixtral, Nemotron) at 4\,k (\S\ref{sec:results}).
Five-seed means and CIs are in \S\ref{sec:seeds} (e.g.\ Mistral-7B
$+3.52\%$ single-seed $\to +2.90\%$ five-seed mean).}
\label{tab:sweep}
\begin{tabular}{lrrrrrrr}
\toprule
Model & Release & Params & $n_\text{layers}$ & $d_\text{model}$ & $k{=}25\%$ & $k{=}75\%$ & \textbf{$k{=}100\%$} \\
\midrule
\textbf{Qwen3-32B} & Apr\,2025 & 32\,B & 64 & 5120 & $+0.27\%$ & $+0.78\%$ & $\boldsymbol{+2.62\%}$ \\
\textbf{Mistral-7B-v0.3} & May\,2024 & 7\,B & 32 & 4096 & $+0.70\%$ & $+1.31\%$ & $\boldsymbol{+3.52\%}$ \\
\textbf{Qwen3-8B} & Apr\,2025 & 8\,B & 36 & 4096 & $-0.57\%$ & $+1.12\%$ & $\boldsymbol{+4.35\%}$ \\
\textbf{Mistral-Small-24B} & Jan\,2025 & 24\,B & 40 & 5120 & $+1.03\%$ & $+1.72\%$ & $\boldsymbol{+4.41\%}$ \\
TinyLlama-1.1B & Jan\,2024 & 1.1\,B & 22 & 2048 & $+0.69\%$ & $+3.10\%$ & $+6.06\%$ \\
SmolLM3-3B & Jul\,2025 & 3\,B & 36 & 2048 & $+1.36\%$ & $+3.66\%$ & $+7.13\%$ \\
Qwen3-4B & Apr\,2025 & 4\,B & 36 & 2560 & $-1.44\%$ & $+1.27\%$ & $+7.34\%$ \\
Qwen3-30B-A3B$^\dag$ & Apr\,2025 & 30/3\,B & 48 & 2048 & $+1.59\%$ & $+2.80\%$ & $+13.78\%$ \\
Mixtral-8x22B$^\dag$ & Apr\,2024 & 141/39\,B & 56 & 6144 & $+10.31\%$ & $+21.94\%$ & $+1800\%$ \\
Mistral-Nemo-12B & Jul\,2024 & 12\,B & 40 & 5120 & $+2.01\%$ & $+4.96\%$ & $+25.45\%$ \\
Granite-3.1-8B & Nov\,2024 & 8\,B & 40 & 4096 & $+4.81\%$ & $+13.94\%$ & $+24.64\%$ \\
Llama-3.1-Nemotron-70B$^\ddag$ & Oct\,2024 & 70\,B & 80 & 8192 & $+14.79\%$ & $+29.41\%$ & $+40.28\%$ \\
\midrule
\emph{GPT-2 (124\,M, 2019, for ref.)} & 2019 & 0.12\,B & 12 & 768 & $+73\%$ & n/a & $+5538\%$ \\
\bottomrule
\end{tabular}
\end{table}

\paragraph{Primary dense-model result.} Qwen3-32B
(32\,B, 64 layers, 4$\times$A40 tensor-sharded) takes $+2.62\%$ with all
attention layers wrapped. This establishes that the behavioral \fcdc noise
model can be tolerated by a modern dense LLM at substantially larger scale than
the GPT-2-class demonstrations in prior analog-attention work \cite{leroux2025}.
Mixtral-8x22B at $k{=}75\%$ (42 of 56 layers analog, 141\,B parameters,
8$\times$A40 sharded) is reported as a partial-layer stress test rather than as
a full all-layer deployment result.

\subsection{Empirical noise-tolerance patterns}\label{sec:laws}

The 12-model sweep (plus the GPT-2 124\,M reference row), with per-architecture controls, supports three empirical
patterns.

\paragraph{Scale improves noise tolerance within a family.} Within the Qwen3 family, $4$B $\to$ $8$B $\to$
$32$B descends monotonically ($+7.34\% \to +4.35\% \to +2.62\%$).
Mistral-7B $\to$ Mistral-Small-24B is mildly non-monotone
($+3.52\% \to +4.41\%$), attributed to tokenizer and pretraining-corpus
differences between the two releases rather than to a counter-example of
the scaling pattern. More parameters per dimension provide more
redundancy to absorb additive Gaussian noise.

\paragraph{MoE attention routing is sensitive at $k{=}100\%$.} Both MoE
models, Qwen3-30B-A3B and Mixtral-8x22B, tolerate $k\!\leq\!75\%$ analog
($+2.80\%$ and $+21.94\%$ respectively) but degrade sharply at $k{=}100\%$
($+13.78\%$ and $+1800\%$). The router amplifies attention noise into expert
selection errors, which compound. The LoRA QAT method of \S\ref{sec:qat}
provides a natural adaptation path; MoE-routed serving remains future work.

\paragraph{Instruction tuning reduces noise headroom.}
Llama-3.1-Nemotron-70B-Instruct has reference PPL 2.498 (vs 7.3 for base
Qwen3-32B); this leaves limited margin. The $+40.28\%$ hit reflects this: a
peaked output distribution is destabilized faster by noise. For
deployment, instruction tuning should therefore follow analog
substitution (or QAT) rather than precede it.

\subsection{Noise tolerance curve}\label{sec:noisecurve}

A noise sweep on TinyLlama-1.1B (8\,k tokens, all 22 layers analog) shows a clear phase transition:
PPL remains usable for $\mathrm{nf}\!\leq\!0.025$ and degrades sharply
beyond. The nominal operating point $\mathrm{nf}{=}0.015$ sits in the
stable regime with $\sim 1.7\times$ noise margin to the transition
(Fig.~\ref{fig:noisecurve}):

\begin{figure}[!ht]
\centering
\includegraphics[width=0.78\linewidth]{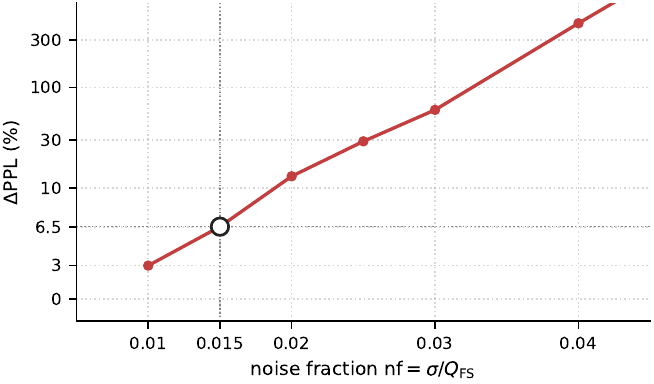}
\caption{TinyLlama WikiText-2 $\Delta$PPL versus per-cell noise fraction
$\mathrm{nf}{=}\sigma/Q_{\mathrm{FS}}$. Symlog $y$ axis (linear below
$10\%$, log above). Hollow marker: \fcdc operating point
($\mathrm{nf}{=}0.015$, $\Delta\mathrm{PPL}{=}+6.5\%$).}
\label{fig:noisecurve}
\end{figure}

\begin{center}
\begin{tabular}{lrrrrrrr}
\toprule
$\mathrm{nf}$ & 0.010 & \textbf{0.015} & 0.020 & 0.025 & 0.030 & 0.040 & 0.060 \\
$\Delta$PPL & $+3.0\%$ & $\boldsymbol{+6.5\%}$ & $+13.0\%$ & $+29.1\%$ & $+59.9\%$ & $+439\%$ & $+21{,}200\%$ \\
\bottomrule
\end{tabular}
\end{center}

\paragraph{Cross-architecture design-space sweep.} To confirm the transition is
not TinyLlama-specific, the three tile operating points were run
$\mathrm{nf}\in\{0.009, 0.015, 0.035\}$ from \S\ref{sec:noisemodel} across
three production LLMs (Qwen3-32B, Mistral-Small-24B-Base, gemma-4-31B). At the
conservative tile $\mathrm{nf}{=}0.009$ all three models stay within $+8\%$ PPL
of FP reference at $k{=}25\%$ analog; at the aggressive tile $\mathrm{nf}{=}0.035$
the transition has been crossed and PPL degrades sharply for every model. The
architectural ranking of which models tolerate analog substitution is
preserved across the design space, indicating that downstream tile-design
trade-offs (area vs noise margin) do not require re-evaluating model
choice.

\section{Noise-Aware LoRA Adaptation}\label{sec:qat}
\FloatBarrier

For older or LayerNorm-based models, and as a path for MoE routing, the
analysis provides a noise-aware LoRA \cite{hu2022lora} quantization-aware
training method. The method trains low-rank adapters (rank 8) on top of
the noisy $q,k,v,o$ projections against the \fcdc noise model. \emph{LoRA
adapters are trained only on the WikiText-2 \cite{merity2017wikitext}
training split; all reported PPL values are on the held-out WikiText-2
test split.} Training runs in $\sim 8$ seconds on one A40 for the
24\,K-parameter k${=}$6 case.

\begin{table}[!ht]
\centering
\small
\caption{GPT-2 $k{=}6$ analog attention: unadapted vs LoRA QAT.}
\label{tab:gpt2lora}
\begin{tabular}{@{}lrr@{}}
\toprule
Configuration & PPL (WikiText-2) & $\Delta$ vs ref \\
\midrule
Reference (no \fcdc) & 32.64 & n/a \\
k${=}$6 unadapted & 52.80 & $+61.77\%$ \\
\textbf{k${=}$6 + LoRA (24\,K params)} & \textbf{33.51} & $\boldsymbol{+2.66\%}$ \\
\bottomrule
\end{tabular}
\end{table}

A 3-phase curriculum (k${=}4 \to 8 \to 12$, 590\,K LoRA params total)
shows that 8 of 12 GPT-2 attention layers can be \fcdc-analog with
a negative PPL hit ($-3.49\%$): the LoRA adapter acts as a mild
regularizer. Only the embedding-adjacent and output-adjacent attention
layers (0, 1, 10, 11) refuse to fine-tune cleanly, a known empirical
property of transformer stacks: these are exactly the layers prior work
flags as asymmetrically important.

\paragraph{LoRA QAT at LLM scale.} The same method scales to the
frontier setting: rank-8 adapters on all $q,k,v,o$ projections
($2.3$\,M parameters on TinyLlama-1.1B, $6.8$\,M on Mistral-7B-v0.3,
${\sim}0.1\%$ of model size) are trained for $300$--$400$ steps on
$120$\,k WikiText-2 \emph{training} tokens with the full
noisy/quantized forward ($\mathrm{nf}{=}0.015$, 8-bit DAC/ADC,
$k{=}100\%$). Because the adapter also confers in-domain adaptation,
the controlled comparison is against the \emph{same} LoRA trained on
the clean model: relative to that control, noise-aware training shrinks
the all-layer analog penalty from $+2.8\%$ to $+0.8\%$ on Mistral-7B
and from $+6.5\%$ to $+2.9\%$ on TinyLlama. Training takes under two
minutes on one A40 for Mistral-7B.

\section{Dithered Resilience to Input PWM Non-Idealities}\label{sec:stride}
\FloatBarrier

Section~\ref{sec:hwtable} identified input-side PWM pulse-width INL/DNL as the
binding non-ideality: because it distorts the row-drive \emph{value} that
multiplies into every term of the charge accumulation, a static $0.5$\,LSB
error already adds $+7.4$\,pp and $\geq\!1$\,LSB collapses the model
($+181.9\%$/$+183.4\%$ on TinyLlama / Mistral-7B), while the non-calibratable
timing jitter is worse still ($+22.0\%$/$+307.3\%$ at $1$\,LSB). Noise-aware
training can recover this, but only by retraining each model, and, as shown
below, that training becomes unstable at $7$\,B scale. We therefore ask a
sharper question: \emph{where} in attention does the input distortion do its
damage, and can it be neutralized in the periphery without retraining?
Protocol throughout this section: TinyLlama and Mistral-7B-v0.3,
$1$--$2$\,k WikiText-2 test tokens (shorter than the main sweep's
$4$--$8$\,k, which is why the vanilla collapse magnitudes quoted here differ
from the $4$\,k-token values of \S\ref{sec:hwtable}), nominal
$\mathrm{nf}{=}0.015$, $k{=}100\%$, single seed.

\paragraph{The value projection is the fragile site.} Sweeping the per-projection
read effort by leave-one-out (each projection module individually returned to
the undefended path while all others stay protected, then averaged by
projection type) localizes the fragility sharply onto the \emph{value}
projection (Table~\ref{tab:vfrag}). On TinyLlama, returning a single $q$, $k$,
or $o$ projection is free (the $\Delta$PPL is within noise of zero, even
slightly negative), while a single $v$ projection costs $+2.5$\,pp on average;
Mistral shows the same ordering with a softer margin.
The mechanism is that $q$ and $k$ set only the attention \emph{weights}, and the
softmax normalization is forgiving of small score perturbations, whereas $v$
carries the \emph{content} that is aggregated, so an input distortion there
propagates directly to the layer output. This refines the \S\ref{sec:qat}
observation that output-adjacent layers are asymmetrically important into a
per-projection statement: under analog input noise, protect the value path.

\begin{table}[!ht]
\centering
\small
\caption{Per-projection fragility: mean $\Delta$PPL (pp) from returning a
single projection module of the given type to the undefended path
(leave-one-out, averaged over layers), at $1$\,LSB INL. The value projection
dominates.}
\label{tab:vfrag}
\begin{tabular}{@{}lrrrr@{}}
\toprule
Model & $q$ & $k$ & $v$ & $o$ \\
\midrule
TinyLlama-1.1B & $-0.2$ & $-0.2$ & $\mathbf{+2.5}$ & $-0.2$ \\
Mistral-7B-v0.3 & $+0.6$ & $+0.6$ & $\mathbf{+1.6}$ & $+0.6$ \\
\bottomrule
\end{tabular}
\end{table}

\paragraph{Input dithering.} The defense is a periphery-side encoding we call
\emph{input dithering}: a transformer-scale instance of the dynamic-element-matching
/ data-weighted-averaging idea long used to linearize DACs \cite{baird1995dwa}.
The static INL is a fixed, repeatable function of the $8$-bit code: a given input
value always rounds to the same code and always picks up the same
$\mathrm{INL}[\text{code}]$, a signal-correlated bias the read accumulation
cannot average away. Dithering breaks that determinism: each read adds a known
common-mode integer code offset $c$ (drawn per token from a tile-resident
pseudorandom stream), and the periphery subtracts $c$ after digitization. Over
$K$ sub-reads the recovered value is $v + \tfrac{1}{K}\sum_j
\mathrm{INL}[\text{code}+c_j]$; because each $c_j$ lands the value on a
\emph{different} INL bin, the residual is zero-mean and its standard deviation
falls as $1/\sqrt{K}$, so it now averages like the benign additive read noise the
model already tolerates at $\mathrm{nf}{=}0.015$. The offset subtraction is a
single $c_j\!\cdot\!\mathrm{colsum}(W)$ scalar per read, not a second
matmul, and is \emph{exact} at $\mathrm{INL}{=}0$: the dithered path reproduces
the undithered $8$-bit baseline to machine precision. The identical mechanism
averages the random jitter at no extra cost.

\paragraph{Training-free recovery.} With $K{=}16$ sub-reads and no retraining,
dithering pulls the $1$\,LSB INL collapse from $+181.9\%$/$+183.4\%$ back to
$+8.8\%$/$+6.0\%$, within a few pp of the $\mathrm{INL}{=}0$ baseline
($+7.6\%$/$+2.3\%$), and the $1$\,LSB jitter collapse from $+22.0\%$/$+307.3\%$
to $+6.7\%$/$+3.5\%$ (Table~\ref{tab:stride}). Swept against an \emph{ideal}
per-tile INL look-up table (the strongest static-calibration baseline, which
subtracts the exact INL profile), the comparison is two-sided: the ideal LUT
is \emph{better} than dithering on pure static INL (it is exact by construction),
but it does \emph{nothing} for jitter, with LUT and undefended perplexity
identical across the jitter sweep, whereas dithering addresses both. The INL
failure threshold (the LSB at which $\Delta$PPL crosses baseline${+}5\%$) moves
from $0.5$\,LSB undefended to $1.5$\,LSB under dithering, a $3\times$ relaxation
of the delay-line linearity spec, and the jitter threshold never crosses within
the swept range.

\begin{table}[!ht]
\centering
\small
\caption{Training-free dithering ($K{=}16$) vs the undefended path and an ideal
static INL LUT: $\Delta$PPL vs the clean reference at $\mathrm{nf}{=}0.015$,
$k{=}100\%$. The LUT is exact on static INL but cannot touch jitter; dithering
addresses both at $K\times$ read cost.}
\label{tab:stride}
\begin{tabular}{@{}llrrr@{}}
\toprule
Model & Stressor ($1$\,LSB) & undefended & dithered & ideal LUT \\
\midrule
TinyLlama   & static INL    & $+181.9\%$ & $+8.8\%$ & $+7.6\%$ \\
            & timing jitter & $+22.0\%$  & $+6.7\%$ & $+22.0\%$ \\
Mistral-7B  & static INL    & $+183.4\%$ & $+6.0\%$ & $+2.3\%$ \\
            & timing jitter & $+307.3\%$ & $+3.5\%$ & $+307.3\%$ \\
\bottomrule
\end{tabular}
\end{table}

\paragraph{Dithering stabilizes noise-aware training.} Dithering and the
noise-aware LoRA of \S\ref{sec:qat} are complementary, and combining them is
both better and, more importantly at scale, more stable. Training the rank-8
adapter \emph{through} the dithered forward (matched parameter and step budget)
beats the undithered noise-aware LoRA on every condition: on Mistral-7B the
combined-stressor penalty is $+5.9\%$ with dithering versus $+28.7\%$ without,
and on the non-calibratable jitter axis $+4.7\%$ versus $+12.1\%$. The
undithered noise-aware LoRA is moreover unstable at $7$\,B: across repeated
runs through the heavy INL${+}$jitter forward it diverged to
$+10^3$--$10^4\%$ perplexity despite gradient clipping, whereas the dithered
training, which sees a smaller zero-mean residual, converged consistently to
${\sim}+5\%$. Dithering thus removes a practical barrier to noise-aware QAT
at scale.

\paragraph{Cost and targeted allocation.} The recovery is not free: the $K$
sub-reads multiply the per-read tile energy (row drive and ADC conversion,
as evaluated here) by $K$; an analog accumulate-then-convert variant could
hold conversions at one per output, at the cost of $K\times$ integration
headroom, but is not assumed. Uniform $K{=}16$ is impractical, but the
value-projection localization makes it unnecessary: protecting the fragile
projections at high $K$ and the rest at a low floor reaches within $2$\,pp of
uniform-$K{=}16$ quality at $4$--$6\times$ average reads (Mistral $+6.0\%$ at
$5.8\times$; TinyLlama $+10.8\%$ at $4.1\times$). At that point the dithered
tile costs ${\approx}3.8$--$5.6$\,fJ/MAC, still below the 8-b-projected
SC-SRAM comparison point of \S\ref{sec:system}; and dithering is a fallback
for tiles that miss the $<\!0.5$\,LSB linearity spec of \S\ref{sec:hwtable},
not a tax on the nominal operating point. Unlike retraining, it needs no
per-model adaptation.

\paragraph{Summary.} The contribution of this section is the localization of
analog-input fragility to the value projection, and the demonstration that a
training-free, calibration-free periphery primitive neutralizes the binding
PWM non-ideality, including the non-calibratable jitter that static
calibration cannot touch, while stabilizing noise-aware training at LLM
scale.

\section{System-Level Energy Analysis}
\FloatBarrier\label{sec:system}

\subsection{Per-token attention energy vs GPU}

A \fcdc tile is compared to an A40 GPU on the per-token attention
energy of one Mistral-7B-class layer. A decode reads and dots against
$T$ keys and $T$ values, so the per-token tile cost grows linearly in
$T$ from row-drive, column-reset, ADC, and local-accumulation traffic;
we cost it from the per-MAC tile energy rather than from a flat sweep.

\paragraph{Per-token scaling.} A Mistral-like layer at $T{=}1024$ with
32 heads and $d_\mathrm{head}{=}128$ requires
$2{\cdot}T{\cdot}32{\cdot}128\!\approx\!8.4{\times}10^6$
MACs per token across $QK^\top$ and $AV$. Using the per-MAC tile energy
at the 8-bit operating point (\S\ref{sec:hwtable}), the FCDC side is
$7.9{\times}10^{-9}$\,J/token (PWM, $0.94$\,fJ/MAC) or
$3.2{\times}10^{-7}$\,J/token (amplitude V-DAC, $38$\,fJ/MAC).
Against the same analytic A40 baseline of $1.99{\times}10^{-6}$\,J/token,
the per-MAC attention-energy advantage at $T{=}1024$ is
$\sim\!250\times$ (PWM operating point) to $\sim\!6\times$
(amplitude V-DAC). This is a compute-only figure; the
GPU$\leftrightarrow$\fcdc data movement it omits is bounded in
\S\ref{sec:serving} (Table~\ref{tab:io}) and erodes the $250\times$ to
$\sim\!1.7$--$6.6\times$ unless the substrate is co-packaged.

The ratio still grows with $T$, because GPU high-bandwidth-memory (HBM) byte cost scales as
$O(T)$ per generated token (KV-cache transfer) while FCDC's $O(T)$ term
is column-charge rather than HBM; system-level routing, control, and
on-GPU post-processing are not folded into either per-MAC column.

\paragraph{Comparison vs measured analog-IMC silicon.} The appropriate
peripheral-matched comparator for active-MAC energy is not a GPU but
prior analog-IMC silicon. The closest comparators are collected:

\begin{center}
\small
\resizebox{\textwidth}{!}{%
\begin{tabular}{@{}p{0.48\textwidth}rp{0.34\textwidth}@{}}
\toprule
Macro / projection & fJ/MAC (norm.) & notes \\
\midrule
HERMES, 14\,nm PCM-IMC, 1-phase (8b in/out) \cite{legallo2023hermes}
  & $\approx 204$ & 9.76 TOPS/W; full 64-core chip incl. control \\
HERMES, 4-phase high-precision \cite{legallo2023hermes}
  & $\approx 806$ & 2.48 TOPS/W; same chip, precision mode \\
Princeton SC-SRAM, 28\,nm (5b in) \cite{verma2021charge}
  & $\approx 0.17$ & 5796 TOPS/W, 1-b-normalized \\
Princeton SC-SRAM, differential, 2024 \cite{verma2024scsram}
  & $\approx 0.12$ & 8161 TOPS/W, 1-b-normalized; 111.8 TOPS/mm$^2$ \\
\midrule
\textbf{FCDC PWM (this work, op.\ point)} & $\mathbf{0.94}$ & projected tile model, 8-bit \\
\textbf{FCDC V-DAC (this work, alt.)}   & $\mathbf{38}$ & projected, 8-bit amplitude DAC \\
\bottomrule
\end{tabular}
}
\end{center}

\textbf{Normalization (apples-to-apples).} The four comparators use
different precisions and reporting conventions; they are listed
explicitly so the fJ/MAC column can be read on a common basis.
(i) \emph{HERMES} numbers are measured silicon at the full 64-core
chip level (1-phase 8-b in/out and a 4-phase higher-precision mode);
fJ/MAC is quoted at the native 8-b precision, not 1-b-normalized.
(ii) \emph{Princeton SC-SRAM} numbers are measured silicon, but the
headline TOPS/W and the fJ/MAC entry above are \emph{1-b-normalized}
(i.e.\ a 1-b$\times$1-b MAC); at 4-b weights, comparable to the FCDC
rows, the per-MAC energy scales by roughly $4^2{=}16\times$, giving
${\sim}2.7$ and ${\sim}1.9$ fJ/MAC for the 28\,nm and differential
variants respectively.
(iii) The FCDC entries are \emph{projected} from the tile model, not
measured silicon, at the 8-bit operating point.
With those caveats lined up, FCDC's projected PWM tile ($0.94$\,fJ/MAC)
is ${\sim}200\times$ below HERMES and below a 4-b-projected SC-SRAM
($1.9$--$2.7$\,fJ/MAC); at the matched 8-bit precision the gap widens to
${\sim}10\times$ below an 8-b-projected SC-SRAM ($\sim\!8$--$11$\,fJ/MAC),
because PWM row energy is resolution-independent while charge-domain
SC-SRAM scales as ${\sim}2^{2B}$. The amplitude V-DAC variant
($38$\,fJ/MAC at 8-bit) sits ${\sim}5\times$ below HERMES but above
SC-SRAM. The
relevant FCDC differentiator over SC-SRAM is therefore \emph{not} only
active-MAC energy but non-volatility (no refresh), KV-cache residency,
and BEOL stackability. PCM-based HERMES is non-volatile too, but
does not pursue in-cell attention compute. This comparison is the
peripheral-matched active-MAC baseline for the projected tile.

\subsection{Measured A40 baseline (BF16, INT8, INT4)}\label{sec:measured-gpu}

The A40 column above is an analytic FLOPs $+$ HBM-byte model. The
corresponding energy is also measured on real silicon, both for prefill
attention only and for end-to-end autoregressive decode at three
weight precisions: BF16, INT8 (bitsandbytes), and INT4 (\texttt{nf4}).

\paragraph{Attention-only prefill.} A Mistral-7B head configuration
(32 heads, head\_dim$=$128, 8 KV heads, batch 16) is run on
A40, using the NVIDIA Management Library (NVML) on-device energy counter and the same configuration as
the analytic model. $\geq 2$\,J per measurement is integrated to
dominate counter resolution.

\begin{center}
\small
\begin{tabular}{lrrrr}
\toprule
Impl.\ / dtype & $T$ & measured J/prefill-tok & analytic model & measured/model \\
\midrule
SDPA bf16  & 16   & $6.76{\times}10^{-5}$ & $3.72{\times}10^{-8}$ & $1.8{\times}10^{3}$ \\
SDPA bf16  & 64   & $1.73{\times}10^{-5}$ & $1.30{\times}10^{-7}$ & $1.3{\times}10^{2}$ \\
SDPA bf16  & 256  & $2.38{\times}10^{-5}$ & $5.01{\times}10^{-7}$ & $4.8{\times}10^{1}$ \\
SDPA bf16  & 1024 & $3.92{\times}10^{-5}$ & $1.99{\times}10^{-6}$ & $2.0{\times}10^{1}$ \\
\bottomrule
\end{tabular}
\end{center}

The measured A40 energy is $20\!\!-\!\!1800\times$ higher than the
analytic FLOP$+$HBM model, dominated by GPU baseline power that the
analytic model excludes. Idle board power is $\sim\!70$\,W via NVML;
the sustained per-precision averages (BF16 $\sim\!250$--$269$\,W, INT4
$\sim\!168$--$178$\,W) appear in the end-to-end decode table below.
Transient NVML samples during short prefill bursts can exceed the
A40's $300$\,W nominal thermal design power (TDP) because NVML reports instantaneous
board-power estimates rather than time-averaged dissipation; the
energy ratios in the rest of this paper use the $\geq\!2$\,J-integrated
per-token energies in the table below, not the burst-power readings.

\paragraph{End-to-end decode (full model).} A more realistic GPU
baseline is the per-decoded-token wall energy of an autoregressive
Mistral-7B forward pass. This is measured at three precisions, using
NVML with $\geq 2$\,J per measurement target:

\begin{center}
\small
\begin{tabular}{lrrrr}
\toprule
Precision & $T$ & J / decoded token & tokens/s & avg W \\
\midrule
BF16  & 16   & $5.59$ & $44.7$ & $250$ \\
BF16  & 1024 & $5.95$ & $44.1$ & $262$ \\
BF16  & 4096 & $6.45$ & $41.7$ & $269$ \\
INT8 (bnb) & 1024 & $10.50$ & $12.7$ & $134$ \\
INT4 (nf4) & 16   & $4.43$ & $37.8$ & $168$ \\
INT4 (nf4) & 1024 & $4.51$ & $37.8$ & $171$ \\
INT4 (nf4) & 4096 & $4.70$ & $37.8$ & $178$ \\
\bottomrule
\end{tabular}
\end{center}

INT4 is the strongest realistic GPU baseline for a single-user serving
workload: it dominates BF16 by $\sim 1.3\times$ wall energy at $T{=}1024$
while staying within $\sim 15\%$ of BF16 throughput. This measured INT4
number is used as the GPU side of the workload-level comparison
in \S\ref{sec:serving} and as the active-MAC reference baseline going
forward; the earlier analytic BF16 column is retained only for
traceability.

\subsection{Cache energy vs volatile gain cells}\label{sec:cache}

The most important system-level differentiator vs \cite{leroux2025} is
\emph{idle / parked} cache cost. The intrinsic ferroelectric switching
work per cell is
$E_\mathrm{sw}{\sim}V Q_\mathrm{sw}{=}1.2\,\mathrm{V}\cdot 2P_r A
\approx 1.5$\,fJ for a 50\,nm cell with $P_r{=}25\,\mu$C/cm$^2$, but
array-level write energy (write-driver, row/bit-line charging,
write-verify) is $10\!-\!10^3\times$ larger; the analysis therefore reports the
gain-cell crossover as a \emph{range} parameterized by the effective
FCDC write energy. With $E_\mathrm{write}^\mathrm{gc}{=}5{\times}10^{-14}$\,J,
$\tau_\mathrm{refresh}{=}1$\,ms, and head\_dim$=$64, \fcdc beats a parked gain-cell cache after
$\sim\!E_\mathrm{write}^\mathrm{fcdc}/(50\,\mathrm{fJ}\cdot 1\,\mathrm{ms}^{-1})
\sim 2{\times}10^{-5}$ to $2{\times}10^{-3}$\,s of residency depending on whether the
FCDC write is the intrinsic $1$\,fJ or an array-level $100$\,fJ. At 28\,h steady-state the parked-cache advantage is
$\sim\!5{\times}10^7$ (assuming $E_\mathrm{write}^\mathrm{fcdc}\!\sim\!100$\,fJ) to
$\sim\!5{\times}10^9$ (intrinsic $1$\,fJ). For an active cache read once per second, the active read-energy term
dominates over refresh once residency exceeds the read interval; the
FCDC active advantage at 28\,h drops to $9.5\times$ ($\tau{=}1$\,ms) to
$85.7\times$ ($\tau{=}0.1$\,ms).

This is what makes \fcdc valuable specifically for long-context inference,
agentic loops with idle gaps, RAG with persistent caches, and multi-tenant
serving with pinned caches. None of those regimes are served well by a
volatile array.

\subsection{KV-cache capacity and area}\label{sec:area}

A literal mapping of the KV cache onto compute tiles looks alarming. A
Mistral-7B-class 8\,k cache holds
$2{\cdot}n_\mathrm{layer}{\cdot}n_\mathrm{kv}{\cdot}d_\mathrm{head}{\cdot}T
{=}537$\,M analog scalars; at $65{,}536$ cells per tile that is $8192$
tiles, and at the $0.2$--$0.4$\,mm$^2$ peripheral-limited tile size of
\S\ref{sec:tile} this implies $1.6$--$3.3{\times}10^3$\,mm$^2$ of silicon.
That estimate is a real worst case, but it assumes one
\emph{fully-peripheralized compute tile} (128 ADCs, 256 drivers) per
$65{,}536$ \emph{stored} scalars: every key wired to its own ADC and read
in a single shot, i.e.\ maximum parallelism.

Two facts break that assumption. First, cells are tiny and the storage
cost is irreducible but small: at 50\,nm pitch a cell is
$2.5{\times}10^{-9}$\,mm$^2$, so the entire 8\,k cache is $1.34$\,mm$^2$ of
cells. Storage is not the binding term. Second, ADC count is set by
readout \emph{throughput}, not capacity. One decoded token requires
$n_\mathrm{layer}n_\mathrm{q}[\,T+d_\mathrm{head}\lceil T/256\rceil\,]
{\approx}12.6$\,M conversions at 8\,k; a single 128-ADC macro at the 5\,ns
read therefore sustains ${\approx}2035$ tok/s, $54\times$ the measured INT4
GPU decode rate (\S\ref{sec:measured-gpu}). Cells can be banked behind a
shared readout exactly as DRAM and NAND share wordline drivers down a bank,
so the periphery scales with throughput, not with stored capacity.

\begin{table}[htbp]
\centering
\small
\caption{KV-cache area for one Mistral-7B-class model, single-ended.
\emph{Naive}: one full compute tile per $65{,}536$ stored scalars
($0.228$\,mm$^2$/tile, the component sum of \S\ref{sec:hwtable}).
\emph{Banked}: cells plus the readout macros needed to meet the
$37.8$\,tok/s INT4 GPU decode rate. DRAM column stores the same INT8 cache
(volatile, refresh) at $0.16$\,Gbit/mm$^2$.}
\label{tab:area}
\begin{tabular}{@{}lrrrrr@{}}
\toprule
Context $T$ & KV scalars & Cells & Naive & \textbf{Banked} & DRAM (INT8) \\
\midrule
8\,k   & $537$\,M  & $1.34$\,mm$^2$ & $1869$\,mm$^2$  & $\mathbf{1.65}$\,mm$^2$ & $26.8$\,mm$^2$ \\
32\,k  & $2.15$\,G & $5.37$\,mm$^2$ & $7476$\,mm$^2$  & $\mathbf{5.92}$\,mm$^2$ & $107$\,mm$^2$ \\
128\,k & $8.59$\,G & $21.5$\,mm$^2$ & $29906$\,mm$^2$ & $\mathbf{23.0}$\,mm$^2$ & $429$\,mm$^2$ \\
\bottomrule
\end{tabular}
\end{table}

\begin{figure}[htbp]
\centering
\includegraphics[width=\linewidth]{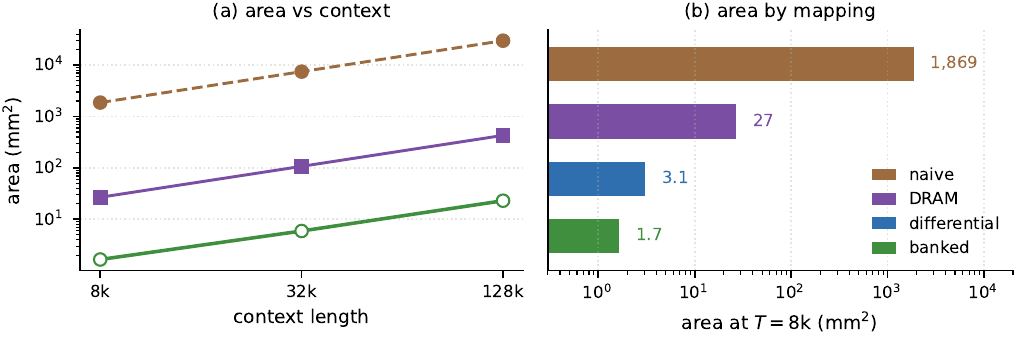}
\caption{KV-cache area for one Mistral-7B-class model. (a) Area versus
context length (colors as in the panel-(b) legend): the banked \fcdc store
(which tracks the cell-array floor) stays far below the naive
one-tile-per-subarray mapping and below DRAM holding the same INT8 cache, at
all context lengths. (b) Area at
$T{=}8$\,k by storage mapping: the banked organization ($1.7$\,mm$^2$) and
its differential variant ($3.1$\,mm$^2$) sit below DRAM ($27$\,mm$^2$) and
${\sim}1100\times$ below the naive mapping ($1869$\,mm$^2$). Throughput is not
the constraint: one 128-ADC macro sustains $54\times$ the INT4 GPU decode
rate (\S\ref{sec:area}).}
\label{fig:area}
\end{figure}

The banked organization (Table~\ref{tab:area}, Fig.~\ref{fig:area}) is
cell-area-dominated: at the single-ended 8-bit operating point it is
$1.65$\,mm$^2$ at 8\,k and $23$\,mm$^2$ at 128\,k, $1131\times$ and
$1300\times$ below the naive mapping, and $16$--$19\times$ smaller than DRAM
holding the same logical cache, while remaining nonvolatile.

The single-ended point assumes one analog scalar per cell, i.e.\ 8 bits in
256 levels, which is aggressive against the $\mathcal{O}(3{-}25)$
grains/cell limit of \S\ref{sec:noisemodel} and would require per-cell
calibration. Table~\ref{tab:area_enc} therefore sweeps the storage encoding,
trading cells per scalar for resolution per cell. Per stored bit the cell
ranges from $312$\,nm$^2$/bit (single-ended 8-bit) to $2500$\,nm$^2$/bit
(grain-conservative 1-bit-per-cell binary), versus ${\sim}6250$\,nm$^2$/bit
for a commodity DRAM array. Even the safest 8-cell binary layout stays
$2.5\times$ denser than DRAM and nonvolatile, so the area advantage does not
rest on the aggressive analog-density assumption; the recommended
differential coding (\S\ref{sec:noisemodel}) sits at $3.1$\,mm$^2$ /
$10\times$ DRAM.

\begin{table}[htbp]
\centering
\small
\caption{Storage-encoding variants at $T{=}8$\,k (Mistral-7B-class, banked).
\emph{Single-ended} and \emph{differential} store 8 bits/cell (calibration
required); \emph{4-/8-cell} split the 8-bit scalar across cells to respect
the grain limit. Density is per stored information bit; DRAM is
${\sim}6250$\,nm$^2$/bit.}
\label{tab:area_enc}
\begin{tabular}{@{}lrrrr@{}}
\toprule
Encoding & cells/scalar & bits/cell & area (mm$^2$) & vs.\ DRAM \\
\midrule
Single-ended    & 1 & 8 & $1.65$  & $20\times$ \\
Differential    & 2 & 8 & $3.08$  & $10\times$ \\
4-cell split    & 4 & 2 & $5.92$  & $5\times$ \\
8-cell binary   & 8 & 1 & $11.62$ & $2.5\times$ \\
\bottomrule
\end{tabular}
\end{table}

\paragraph{A concrete banked organization.} Table~\ref{tab:macro} makes the
banked column concrete with one explicit two-level floorplan for
the 8\,k cache. $B_1$ subarrays share an integration cap and sense amp through
a local column mux; this sharing is bounded not by area but by \emph{charge
division}: the deselected mux junctions add to the bitline capacitance the
$400$\,fF integration node sees, and holding that loss to $\leq\!1\%$
(\S\ref{sec:tile}) caps $B_1$ at $7$ ($C_\mathrm{BL}{=}4.0$\,fF, $0.99\%$
loss). That yields $1171$ integration/sense nodes; the ADCs, which sit
\emph{after} the charge-domain integration and so add no further
charge-division penalty, are shared across all of them for throughput and a
handful suffice ($391{:}1$ sense-to-ADC mux, depth $9$; per-access mux energy
$0.05$\,fJ, $8\%$ of the ADC conversion energy). The resulting breakdown
totals $1.77$\,mm$^2$, with cells and the per-node integration caps (not the
ADCs) dominating.

\begin{table}[htbp]
\centering
\small
\caption{Analytic banked-macro floorplan for the 8\,k Mistral-7B-class cache
(single-ended). Bank depth is charge-division-limited; ADCs are
throughput-provisioned and shared post-sense.}
\label{tab:macro}
\begin{tabular}{@{}lr@{\hskip 2em}lr@{}}
\toprule
Organization & & Area (mm$^2$) & \\
\midrule
Subarrays (256$\times$256)   & $8192$    & cells              & $1.342$ \\
Bank depth $B_1$             & $7$       & integration caps   & $0.234$ \\
Charge-div.\ loss            & $0.99\%$  & sense amps         & $0.094$ \\
Integration/sense nodes      & $1171$    & ADCs               & $0.003$ \\
Shared ADCs (GPU rate)       & $3$       & row drivers        & $0.050$ \\
Sense:ADC mux                & $391{:}1$ & mux + routing      & $0.050$ \\
Mux energy / access          & $0.05$\,fJ & \textbf{total}    & $\mathbf{1.77}$ \\
\bottomrule
\end{tabular}
\end{table}

This is an analytic floorplan, not an extracted layout: the deselected-junction
capacitance, the depth-$9$ mux tree, and the sense-amp area are engineering
estimates (\S\ref{sec:limits}). Within them, the banked area is a concrete
design point rather than an aspiration, and the charge-division bank-depth
limit ($B_1{\le}7$) is the binding organizational constraint, not ADC or
cell area.

\subsection{Long-context serving energy across workloads}\label{sec:serving}

\begin{figure}[tbp]
\centering
\includegraphics[width=\linewidth]{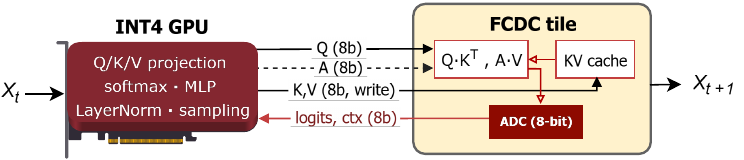}
\caption{Hybrid substrate: INT4 GPU handles $Q/K/V$ projection, softmax,
MLP, and LayerNorm; the \fcdc array stores the KV cache and executes
$Q\!\cdot\!K^\top$ and $A\!\cdot\!V$. 8-bit boundary.}
\label{fig:hybrid}
\end{figure}

\paragraph{Reconciling the two deployment modes.} The LLM sweep in
\S\ref{sec:results} evaluates the \emph{full-substrate} mode in which
the four projections $q,k,v,o$ \emph{and} the two attention
matmuls $Q\!\cdot\!K^\top, A\!\cdot\!V$ are routed through the FCDC
noise model. The serving-energy model below (Fig.~\ref{fig:hybrid})
evaluates the narrower \emph{KV-coprocessor} mode in which only KV
storage and the two attention matmuls live on FCDC while projections,
softmax, MLP, and LayerNorm remain on the GPU. The PPL budgets reported
in \S\ref{sec:results} therefore upper-bound the KV-coprocessor mode,
whose own quality budget is the C5 matmul-only configuration of
\S\ref{sec:e2e}: under $0.5\%$ PPL at $7$--$8$\,B. The harder
full-substrate budget is used throughout because it bounds the
KV-coprocessor mode from above, and because it lets the same noise model
serve both reading regimes.

To turn the cell-level and active-MAC numbers into a serving-level
comparison, per-served-token energy is evaluated across five representative
LLM serving workloads with realistic KV-cache residency patterns. Each
workload is parameterized by the context length $T$, the number of
decoded tokens per session $n_\mathrm{decode}$, the active decode window
$T_\mathrm{active}$, the inter-session residency $T_\mathrm{keep}$, and
the number of sessions per scenario. Six substrates are compared for one
Mistral-7B-class attention layer (grouped-query attention, GQA, $32{\times}q / 8{\times}kv$,
$d_\mathrm{head}{=}128$, 32 layers): measured BF16 and INT4 GPU decode
energy (\S\ref{sec:measured-gpu}, \texttt{nf4} bitsandbytes), measured
SC-SRAM CIM \cite{verma2024scsram}, modeled gain-cell IMC
\cite{leroux2025} with 1\,ms refresh and $5{\times}10^{-14}$\,J/write,
and the two \fcdc tile points (V-DAC and PWM). For the realistic
deployment case the analog substrate handles attention compute and KV
storage while the GPU continues to run MLP, LayerNorm, and sampling at
the chosen precision; this is modeled as ``\fcdc+GPU\,INT4'' with attention
attributed to the analog substrate and the remaining $1-\alpha$ of the
GPU INT4 decode energy attributed to the GPU side ($\alpha{=}0.15$
estimate of attention's share of Mistral-7B decode wall energy at
$T{=}1024$).

\begin{table}[htbp]
\centering
\small
\caption{Per-served-token energy (J) across five workloads. GPU column is
the INT4 GPU baseline; Hybrid is FCDC KV + INT4 GPU decode. Speedup is
the ratio to INT4 GPU; idle power is attributed to the served workload.}
\label{tab:serving}
\begin{tabular}{lrrrrr}
\toprule
Workload & $T$ & $T_\mathrm{keep}$ & GPU (J) & Hybrid (J) & Speedup \\
\midrule
Chat turn          & 8k    & 30\,s   & $25.7$        & $5.76$        & $4.5\times$ \\
Long-context QA    & 32k   & 60\,s   & $21.1$        & $5.65$        & $3.7\times$ \\
RAG with persistent KV & 8k & 1\,h   & $103$         & $5.71$        & $18\times$ \\
Agent loop (10$^2$ tool calls / 8\,h) & 16k & 8\,h & $206$ & $5.82$ & $35\times$ \\
Parked KV (idle 28\,h between turns)  & 8k & 28\,h & $1.10{\times}10^5$ & $84.4$ & $1.3{\times}10^3$ \\
\bottomrule
\end{tabular}
\end{table}

The speedup grows monotonically with $T_\mathrm{keep}$ because GPU idle
power dominates parked sessions, while the \fcdc cache costs zero
refresh and only $\sim\!0.05$\,W chip idle. At the always-on extreme
(28\,h park between turns) the hybrid substrate is dominated by GPU
decode of the rare active token but still wins by three orders of
magnitude because the GPU baseline pays $\sim\!70$\,W idle continuously.
At the chat-turn extreme ($T_\mathrm{keep}{=}30$\,s) the hybrid still
wins by $4{-}5\times$, dominated by the GPU's contribution to MLP and
LayerNorm rather than by attention itself. These ratios use measured
INT4 GPU energy, not a BF16 analytic model, and vary little over
$\alpha\in[0.05, 0.30]$ for the attention fraction.

\paragraph{Serving-side sensitivity bounds.} The serving ratios above
inherit four assumption-sensitive inputs. Each is bounded. (i)
\emph{Attention share} $\alpha$: sweeping $\alpha\in[0.05, 0.30]$
changes the chat-turn ratio by $<\!10\%$ and the agent-loop ratio by
$<\!5\%$. (ii) \emph{FCDC chip idle power}: doubling the assumed $0.05$\,W
idle assumption to $0.1$\,W moves the 28\,h parked ratio from
$1.3{\times}10^3$ to $\sim\!650{\times}$; doubling again to $0.2$\,W
gives $\sim\!325{\times}$. The parked-cache advantage survives
an order of magnitude in chip-idle estimate. (iii) \emph{Effective
write energy}: the cache-energy model already sweeps
$E_\mathrm{write}^\mathrm{fcdc}\in[1\,\mathrm{fJ},\,100\,\mathrm{fJ}]$
(\S\ref{sec:cache}); the dominant active-decode ratio is unaffected
because writes happen at prefill, not at every decode step. (iv)
\emph{Optimized GPU baselines} (vLLM batched, CPU+NVMe parked KV,
power-gated GPU): see Table~\ref{tab:serving_baselines} for the
explicit robustness check. The headline ratios shrink substantially
under these alternatives at short residency, but the parked-cache
advantage survives by $40\!-\!130\times$ and the agent-loop advantage by
$1.9\!-\!4.7\times$ (Table~\ref{tab:serving_baselines}). The qualitative
ordering (parked $\gg$ agent $\gg$ RAG $\gg$ QA $\gg$ chat) is robust
across the entire bound.

\paragraph{Hybrid data-movement (IO) energy.} The active-MAC accounting in
\S\ref{sec:system} excludes GPU$\leftrightarrow$\fcdc traffic. In
KV-coprocessor mode the softmax stays on the GPU, so each decoded token
round-trips the attention scores off the substrate: per layer it moves the
query ($n_\mathrm{q}d_\mathrm{head}$), the scores
($n_\mathrm{q}T$, \fcdc$\to$GPU), the softmax weights ($n_\mathrm{q}T$,
GPU$\to$\fcdc) and the output ($n_\mathrm{q}d_\mathrm{head}$), i.e.\
$2n_\mathrm{layer}n_\mathrm{q}(d_\mathrm{head}{+}T)$ values/token
($17$\,M at 8\,k). At the 8-bit boundary and standard interconnect energies
this leg is bounded as follows.

\begin{table}[htbp]
\centering
\small
\caption{Hybrid IO energy. \emph{Left}: per-layer score round-trip at
$T{=}1024$ against the compute-only $253\times$ active-MAC advantage
(\S\ref{sec:system}). \emph{Right}: full-model IO as a fraction of the
per-served-token serving total (Table~\ref{tab:serving}).}
\label{tab:io}
\begin{tabular}{@{}lrrr@{\hskip 1.5em}r@{}}
\toprule
Placement & pJ/bit & \fcdc{+}IO & vs.\ GPU & IO \% of serving \\
\midrule
compute only        & --   & $0.008$\,\textmu J & $253\times$ & -- \\
on-package (2.5D)   & $0.5$ & $0.30$\,\textmu J & $6.6\times$ & $0.001\%$ \\
in-package          & $2$   & $1.19$\,\textmu J & $1.7\times$ & $0.005\%$ \\
board (SerDes)      & $5$   & $2.96$\,\textmu J & $0.67\times$ & $0.012\%$ \\
off-board (PCIe)    & $15$  & $8.86$\,\textmu J & $0.22\times$ & $0.068\%$ \\
\bottomrule
\end{tabular}
\end{table}

Two conclusions follow (Table~\ref{tab:io}). The serving ratios are
IO-robust: data movement is $\leq\!0.07\%$ of the per-served-token total at
\emph{any} placement, because those totals are dominated by GPU idle and the
GPU's own MLP/LayerNorm, not by attention traffic, so the long-residency
headline of Table~\ref{tab:serving} is unaffected. The compute-only
$253\times$ active-MAC figure, by contrast, is not a deployable number: the
score round-trip alone exceeds the analytic GPU attention energy once the
link is board-level, so the advantage holds only for an on- or in-package
\fcdc ($6.6\times$ / $1.7\times$) and collapses below $1\times$ off-package.
Co-packaging is therefore required to retain any active-path win, and moving
the softmax onto the substrate (eliminating the score round-trip
entirely) is the clean fix and a natural direction for future work.

\paragraph{Robustness against optimized GPU serving strategies.}
The single-user INT4 GPU baseline above is the standard reference point
but not the most aggressive deployment. The same five
workloads are run under three additional GPU-side strategies, reporting how
much each erodes the FCDC+INT4 hybrid headline. All four baselines
share the measured INT4 GPU decode wall energy
(\S\ref{sec:measured-gpu}); they differ only in how the residency-time
idle power is paid. The strategies are: \textbf{G1 vLLM/PagedAttention}
(idle GPU power divided across $B{=}32$ concurrent sessions);
\textbf{G2 CPU+NVMe park} (KV cache offloaded to NVMe during
$T_\mathrm{keep}$, GPU power-gated to $5$\,W, plus
$\mathrm{KV\_bytes}/3\,\mathrm{GB/s}$ reload latency on reactivation);
\textbf{G3 power-gate} (GPU suspended to $5$\,W during $T_\mathrm{keep}$
with HBM-resident KV preserved and a $1.5$\,s wake-up at full idle).

\begin{table}[htbp]
\centering
\small
\caption{Per-served-token energy ratio of FCDC+INT4 hybrid to four
GPU-side strategies. \textbf{G0}: single-user (paper headline);
\textbf{G1}: vLLM at $B{=}32$; \textbf{G2}: CPU+NVMe parked KV with
$5$\,W power-gated GPU; \textbf{G3}: power-gated GPU with HBM-resident
KV. G1--G3 are analytic models tuned to documented operating points
(not measured on this cluster, \S\ref{sec:limits}); G0 uses the measured
INT4 decode energy of \S\ref{sec:measured-gpu}. Ratios $>1$ mean the hybrid
wins; $<1$ means the GPU strategy is cheaper.}
\label{tab:serving_baselines}
\begin{tabular}{lrrrr}
\toprule
Workload & G0 single-user & G1 vLLM $B{=}32$ & G2 CPU+NVMe & G3 power-gate \\
\midrule
Chat turn          & $4.5\times$        & $0.93\times$        & $1.84\times$        & $1.56\times$ \\
Long-context QA    & $3.7\times$        & $0.92\times$        & $1.69\times$        & $1.40\times$ \\
RAG (1\,h keep)    & $18\times$         & $1.36\times$        & $2.96\times$        & $2.35\times$ \\
Agent loop (8\,h)  & $35\times$         & $1.89\times$        & $4.69\times$        & $3.57\times$ \\
Parked KV (28\,h)  & $1.3{\times}10^3$  & $41\times$          & $1.3{\times}10^2$   & $93\times$ \\
\bottomrule
\end{tabular}
\end{table}

The $4\!-\!10^3\times$ headline ratio is
\emph{not} robust across all workloads. For short-residency chat and
QA, the FCDC hybrid is matched-to-slightly-worse than a well-batched
vLLM deployment ($0.92\!-\!0.93\times$); the advantage in those rows is
a deployment-mode-dependent artifact, not an intrinsic device-physics
win. The genuine, robust regime is the long-residency case: against
these idle-amortization strategies the hybrid still wins by
$\geq\!1.36\times$ on RAG, $\geq\!1.89\times$ on agent loops, and
$\geq\!41\times$ on parked sessions. The FCDC advantage is therefore
correctly framed as a long-lived-KV proposition, not a generic serving
win.

This is the strongest system-level positioning for \fcdc: the substrate
is most useful when the KV cache is long-lived (RAG, agents, multi-tenant
serving), exactly where GPU baselines are weakest because of idle
amortization, and exactly where volatile gain cells \cite{leroux2025}
are weakest because of refresh.

\begin{figure}[htbp]
\centering
\includegraphics[width=\linewidth]{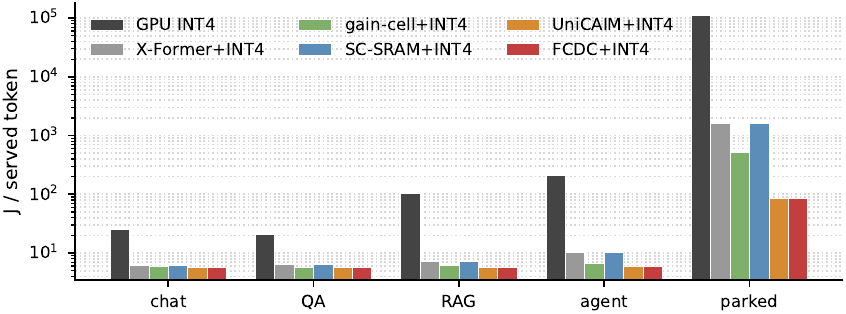}
\caption{Per-served-token energy (J, log scale) for INT4 GPU and five
attention co-processors paired with INT4 GPU, across five long-context
workloads.}
\label{fig:serving}
\end{figure}

\paragraph{Comparison vs FeFET CAM and sparse-attention ASIC baselines.}
The same simulator is extended with two recent published-design baselines
on the same five workloads: a FeFET-based CAM/CIM attention substrate
in the spirit of UniCAIM \cite{unicaim2025} (nonvolatile FeFET array,
top-$k$ similarity-based pruning, $25\%$ effective active fraction,
$0.5$\,fJ/MAC) and a sparse-attention ASIC in the spirit of Sanger /
X-Former (SRAM-resident KV, $10\%$ active fraction, $0.08$\,fJ/MAC).
All three substrates are routed as attention co-processors with INT4
GPU running the rest of decode. The full speedup ratios vs INT4 GPU
are:

\begin{center}
\small
\begin{tabular}{lrrr}
\toprule
Workload & UniCAIM + INT4 & X-Former + INT4 & \fcdc-PWM + INT4 \\
\midrule
Chat turn            & $4.5\times$        & $4.2\times$       & $4.5\times$        \\
Long-context QA      & $3.7\times$        & $3.4\times$       & $3.7\times$        \\
RAG (1\,h keep)      & $18\times$         & $15\times$        & $18\times$         \\
Agent loop (8\,h)    & $35\times$         & $20\times$        & $35\times$         \\
Parked KV (28\,h)    & $1.3{\times}10^3$  & $68\times$        & $1.3{\times}10^3$  \\
\bottomrule
\end{tabular}
\end{center}

\fcdc matches UniCAIM at the workload level on all five points: both
are nonvolatile FE substrates and the chat/QA totals are GPU-other
dominated rather than attention-substrate dominated, so the substrate
type is a wash on short residency. The differentiation appears at
high $T_\mathrm{keep}$: \fcdc and UniCAIM both win the parked workload
by three orders of magnitude, while the SRAM-resident X-Former
baseline pays SRAM leakage on the same long residency and is held to
$68\times$. The contribution claim is therefore not that capacitive
HZO beats UniCAIM, but that the capacitor-based design point reaches
parity with the FeFET CIM state of the art on the long-residency axis
without inheriting FeFET endurance/drift constraints, while remaining
a substantially smaller per-cell device than a 1T1C ferroelectric
trench or 3D-vertical stack.

\section{Limitations}\label{sec:limits}

\paragraph{Validation scope.} No fabricated \fcdc array exists in this work. All device-level numbers come from calibrated multi-tool simulation (\S\ref{sec:crossval}), and a tape-out is required to replace the assumed device-variability distribution with measured statistics. The \textsc{ngspice} Monte-Carlo harness confirms the implementation of the analytic noise model; it is not transistor-level validation of the read switch, sense amplifier, ADC, line coupling, or FE compact model. Similarly, the measured A40 energy in \S\ref{sec:measured-gpu} is not a measured-vs-measured chip comparison; such a comparison requires fabricated \fcdc silicon.

\paragraph{Evaluation scope.} WikiText-2 perplexity is not a complete
downstream-task evaluation. The zero-shot study in \S\ref{sec:lmeval}
covers HellaSwag, ARC, LAMBADA, the full 57-subject MMLU, GSM8K (on the
instruct variant; the base model scores $0$ zero-shot and provides no
separating signal), and a needle-in-a-haystack retrieval test; task
coverage ends there. The longest PPL replication is 128\,k tokens on
Mistral-7B and 32\,k on TinyLlama (\S\ref{sec:longctx}); cache-energy
arguments continue to improve at longer contexts, but 128\,k and above
have not been measured here. The dithering study of \S\ref{sec:stride}
is single-seed on two models. MoE routing remains unresolved at
$k{=}100\%$: the LoRA QAT method works for dense models (GPT-2 through
$k{=}8$; TinyLlama and Mistral-7B at $k{=}100\%$, \S\ref{sec:qat}), but
it has not been trained for an MoE router. The energy comparison uses
measured INT4 GPU decode energy (\S\ref{sec:measured-gpu});
higher-precision matrix engines or fused INT4 kernels could narrow the
gap on the active-MAC axis but not on the idle/parked-cache axis that
drives \S\ref{sec:serving}.

\paragraph{Architecture coverage.} The Llama, Mistral, Qwen, Mixtral, Granite, SmolLM, and Nemotron rows in Table~\ref{tab:sweep} use separate $\mathrm{q\_proj}/\mathrm{k\_proj}/\mathrm{v\_proj}$ modules. Fused-QKV architectures (e.g.\ Phi-3, Phi-4, gemma-4) require three independent tiles per layer and are out of scope here; gemma-4-31B was included only via separate-projection unrolling and remains within $\pm 5$\,pp of the noise-sweep prediction at each $k$.

\paragraph{Device-physics constraints.} NC gain is the least mature ingredient. The 1-D Landau analysis in \S\ref{sec:noisemodel} shows that $|A_v|{=}2.5$ requires $C_s/|C_\mathrm{FE}|\!\approx\!0.71$ (stable for $C_s{<}|C_\mathrm{FE}|$, with an inverted read), a matching condition that swings the gain across $1.5$--$8.4\times$ under $\pm 20\%$ variation in $|C_\mathrm{FE}|$ and therefore requires per-tile trim. A measured non-hysteretic compact-model stack is required before NC gain can be promoted from an exploratory read-path option to a fixed design constant. The Merz/NLS read-disturb bound holds only for $E_\mathrm{eff}{\le}0.20$\,MV/cm on the storage HZO; an NC field enhancement would require a separate field-decoupled stack. The 5\% capacitance-mismatch figure is a post-calibration or differential-coding target, not a raw-cell guarantee (\S\ref{sec:noisemodel}). Finally, the $10^{16}$ endurance and $\geq\!10$\,year retention values are literature anchors from specific 3D 1T-nC-1T stacks \cite{xu2025}, not transferable defaults for the planar 50\,nm model.

\paragraph{System-accounting constraints.} Cell pitch does not by itself produce a tile-area advantage. A 256$\times$256 cell array occupies only $\sim\!1.64{\times}10^{-4}$\,mm$^2$, but a realistic tile is expected to be $0.1\!-\!1$\,mm$^2$ after DACs, ADCs, row drivers, and control logic are floorplanned. The full-cache area therefore depends on how aggressively the readout periphery is banked: \S\ref{sec:area} bounds it between a naive one-tile-per-subarray mapping ($1.6\!-\!3.3{\times}10^3$\,mm$^2$ at 8\,k) and a throughput-banked organization ($\sim\!1.7$\,mm$^2$), the latter contingent on shared charge-domain decode that a tape-out must demonstrate. Single-pulse multi-level analog write is also optimistic: HZO switching is nucleation-limited with broad switching-time distributions, local field inhomogeneity, and charge-injection effects \cite{kondratyukKineticsHZO2022}. Practical deployment should assume write-verify or program-and-calibrate loops: a Merz/NLS Monte-Carlo of program-and-verify (the same kinetics as the read-disturb bound) converges in $6$--$12$ pulses (median; $13$--$18\times$ single-pulse energy) for reachable targets, inside the $10\!-\!10^3\times$ effective write-energy envelope the cache-energy model already sweeps, while the per-cell level count remains grain-limited, consistent with the differential / multi-cell encoding assumption. The peripheral-matched active-MAC baseline is switched-capacitor SRAM CIM, not a GPU; the relevant \fcdc differentiator over that baseline is nonvolatility, no refresh, and KV-cache residency rather than raw active-MAC energy.


\paragraph{Serving-baseline coverage.}
vLLM throughput and NVMe latency were not measured on this cluster; the
G1/G2/G3 columns of Table~\ref{tab:serving_baselines} are analytic
models tuned to documented operating points. Batched decoding on the
FCDC side is out of scope for this paper.

\section{Conclusion}

This paper evaluates HZO ferroelectric capacitors as a nonvolatile
charge-domain substrate for transformer attention. The modeled \fcdc tile
stores analog state in a ferroelectric capacitor, performs local
charge-domain VMM, and is evaluated with device-level noise and peripheral
energy accounting rather than with an idealized array model. Across
\textsc{ngspice}, \textsc{CrossSim}, \textsc{FiPy}, and \textsc{NeuroSim}, the
same analytic tile model gives consistent active-MAC energy estimates.

At the model level, dense LLMs with separate $q,k,v,o$ projections tolerate
the nominal noise point with small perplexity changes: $+2.6\%$ on
Qwen3-32B and, on Mistral-7B-v0.3 (WikiText-2), $+3.5\%$ single-seed at 8\,k
with a $+2.9\%$ five-seed mean ($\pm 0.33$\,pp, \S\ref{sec:seeds}). A LoRA QAT method recovers most of the
degradation in the harder GPT-2 setting and, at LLM scale, shrinks the
all-layer analog penalty to $+0.8\%$ on Mistral-7B relative to an
identically fine-tuned digital baseline. A per-projection analysis
localizes analog-input fragility to the value projection, and
periphery-side input dithering neutralizes the binding PWM input
non-ideality, including the timing jitter that static calibration cannot
touch, without retraining, while stabilizing noise-aware QAT at 7\,B
scale (\S\ref{sec:stride}).

At the system level, the per-token scaling at the 8-bit
operating point projects $\sim\!250\times$ (PWM) to $\sim\!6\times$
(amplitude V-DAC) lower active attention-MAC energy than an analytic A40
baseline at $T{=}1024$. Relative to measured switched-capacitor SRAM CIM,
the PWM tile is at or below it on active-MAC energy, and the durable
advantage is nonvolatility, absence of refresh, and KV-cache residency.
\fcdc is thus a simulation-backed design point for nonvolatile
charge-domain attention; fabrication, longer-context evaluation, and
broader downstream validation are the natural next steps.

\section*{Code availability}

Code is publicly available at \url{https://github.com/faris-agour/FCDC}.
The release contains the LLM noise-injection harness (with Hugging~Face
revision pins, dataset hashes, and per-experiment seed lists), the
analytic tile-energy model with its four cross-implementation drivers
(\textsc{ngspice}, \textsc{CrossSim}, \textsc{FiPy}, \textsc{NeuroSim}),
the NVML measurement scripts for the A40 baselines
(\S\ref{sec:measured-gpu}), and the serving-workload simulator behind
Tables~\ref{tab:serving} and \ref{tab:serving_baselines} with all
baseline settings in a single configuration file. The supplementary
material maps every reported table and figure to its source script and
logged measurement.

\bibliographystyle{plain}
\bibliography{refs}

\appendix
\section{Additional validation details}\label{app:scope}

This appendix collects supplementary validation checks and scope notes
for the reported results.

\paragraph{Physical validation scope.}
The device model is
anchored to measured HZO parameters from the literature (\S\ref{sec:cell});
the thickness-sensitivity sweep stays within $1.4\times$ of the nominal
MAC-energy estimate under $\pm 10\%$ variation (Fig.~\ref{fig:thickness});
and the system-level serving conclusions in \S\ref{sec:serving} are
stress-tested against three optimized GPU baselines (vLLM batching,
CPU+NVMe park, power-gating; Table~\ref{tab:serving_baselines}) and a
sweep of the attention-share $\alpha$, FCDC idle-power, and effective
write-energy inputs (\S\ref{sec:serving}). The remaining uncertainty is
physical implementation risk.

\paragraph{Downstream-task coverage.}
The main quality metric is WikiText-2 perplexity, but \S\ref{sec:lmeval}
also reports the full 57-subject MMLU (14\,k zero-shot questions) and
the full 1\,319-question GSM8K test split for TinyLlama-1.1B-Chat-v1.0
and Mistral-7B-v0.3 (MMLU on the base model; GSM8K on Mistral-7B-Instruct-v0.3
as the base model scores $0$ zero-shot), plus the needle-retrieval test
in \S\ref{sec:lmeval}. Across those runs, the FCDC variant stays within
$\leq 5\%$ relative of the digital baseline; GSM8K is within per-side
stderr ($-0.23$\,pp), while MMLU on Mistral-7B drops $-1.6$\,pp absolute
($-2.7\%$ relative), outside the per-side binomial stderr ($\pm 0.4$\,pp)
but inside the $+3\!-\!7\%$ PPL envelope.

\paragraph{Seed sensitivity.}
The multi-seed study in \S\ref{sec:seeds} reports $n{=}5$ seeds for
short-context Mistral-7B with $\sigma{=}0.33$\,pp on the FCDC delta. The
32\,k-token replication is tighter: mean delta $+2.81\%$, std
$0.11\%$, range $[+2.65\%, +2.96\%]$. The headline therefore does not
depend on a single random seed.

\begin{figure}[tbp]
\centering
\includegraphics[width=0.72\linewidth]{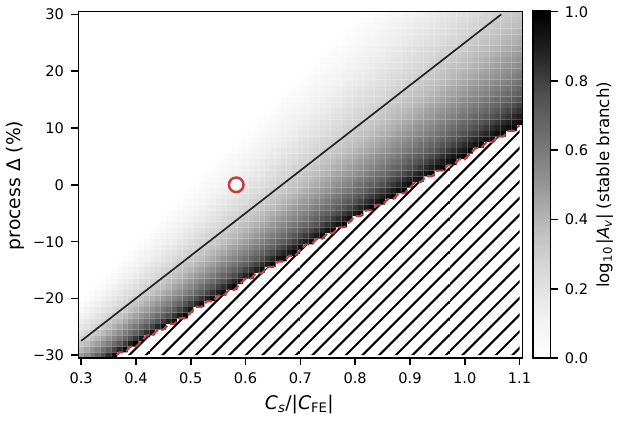}
\caption{NC voltage-gain magnitude over $C_s/|C_{\mathrm{FE}}|$ and
$\pm 30\%$ FE-film process window. Greyscale: $\log_{10}|A_v|$ on the
stable (capacitance-matched) branch, clipped at 10; hatched: runaway
region. Dashed: stability boundary (stable to the left). Solid: design
iso-line $|A_v|{=}2$. Marker: conservative design point
$C_s/|C_{\mathrm{FE}}|{=}0.58$ ($|A_v|{\approx}1.4$) at nominal
process.}
\label{fig:ncstab}
\end{figure}

\paragraph{NC-gain stability.}
The negative-capacitance stability boundary is explicitly swept. The
series stack is stable for $C_s/|C_\mathrm{FE}|{<}1$ (i.e.\
$C_s{<}|C_\mathrm{FE}|$, the capacitance-matching condition); on that
branch the read is inverted and $|A_v|{=}r/(1{-}r)$ rises from below
unity toward the $r{\to}1$ matching pole as the ratio grows. A
Landau--Khalatnikov transient of the read stack confirms the branch
assignment dynamically (the complementary branch runs away onto the
remanent state) and settles to $1\%$ within the $5$\,ns read for
viscosities $\rho\lesssim 1.6\,\Omega\cdot$m at the nominal point. The
body-text nominal point is $|A_v|{=}2.5$ at
$C_s/|C_\mathrm{FE}|{\approx}0.71$ (\S\ref{sec:cell}); under a
$\pm 20\%$ shift in $|C_\mathrm{FE}|$ the gain moves across
$1.5$--$8.4\times$, reaching the matching pole only if
$|C_\mathrm{FE}|$ shrinks ${\sim}29\%$. A more conservative,
larger-margin design point is $C_s/|C_\mathrm{FE}|{=}0.58$ with
$|A_v|{\approx}1.4$, plotted in Fig.~\ref{fig:ncstab}; it remains on
the stable branch across the full $\pm 30\%$ process window (gain
$0.8$--$5.0\times$). As established in \S\ref{sec:noisemodel}, the
headline $\mathrm{nf}{=}0.015$ holds even if NC gain is removed
entirely.

\paragraph{Noise-model coverage.}
The main sweep uses calibrated additive noise. The noise study also
includes per-tile $1/f$ drift, ideal 4/8/10-bit SAR ADC quantization
(no measured INL/DNL, since no ADC is fabricated), and five thickness
corners. Above $\mathrm{nf}{=}0.005$, the aggregate projection-only
frontier is dominated by random Gaussian noise. Below the operating
point the degradation falls off steeply (TinyLlama drops from $+6.5\%$
at $\mathrm{nf}{=}0.015$ to $+3.0\%$ at $\mathrm{nf}{=}0.010$,
\S\ref{sec:noisecurve}), though not uniformly within $\pm 1$\,pp.

\paragraph{MoE routing sensitivity.}
MoE models remain a limited case at $k{=}100\%$. Both MoE models in the
sweep degrade sharply at full analog substitution, so the recommended
deployment point is $k{\leq}75\%$, which preserves most of the analog
energy advantage without destabilizing expert routing. Recovering
$k{=}100\%$ for MoE models requires a router-aware LoRA QAT method and is
left to future work.

\paragraph{Analog-IMC comparison set.}
The system comparison uses the four most relevant published analog-IMC
and attention-co-processor reference points: HERMES PCM-IMC
\cite{legallo2023hermes} and Princeton switched-capacitor SRAM CIM
\cite{verma2021charge,verma2024scsram} as peripheral-matched active-MAC
baselines (\S\ref{sec:system}); the Leroux gain-cell analog-attention
macro \cite{leroux2025} and the UniCAIM FeFET CAM/CIM substrate
\cite{unicaim2025} as long-context attention baselines; and an X-Former-style
sparse-attention ASIC as a SRAM-resident comparator (\S\ref{sec:serving}).
Recent non-archival announcements are excluded when the public data are
insufficient to reproduce the reported operating point.

\paragraph{Analog-tile accounting.}
The hybrid serving model uses the analog-tile costs derived in
\S\ref{sec:hwtable} at the 8-bit operating point: $0.94$\,fJ/MAC total
tile energy in the PWM configuration (an 8-bit amplitude V-DAC is
$38$\,fJ/MAC), with an 8-bit SAR ADC contributing $0.625$\,fJ/MAC and
the PWM DAC contributing $0.31$\,fJ/MAC at the tile level. The
system-level serving advantage in \S\ref{sec:serving} is driven primarily
by the parked-cache idle term on the GPU side, not by assuming zero-cost
analog compute.


\end{document}